\documentclass[draftcls,onecolumn]{IEEEtran}
\usepackage{amsmath,epsfig,amssymb,graphicx,subcaption,caption}
\begin{document}

\newcommand{\Mythetat}{\tilde{\theta}}
\newcommand{\Mygammazero}{\gamma_{0}}
\newcommand{\MyGammazero}{\Gamma_{0}}
\newcommand{\MyGammaK}{\Gamma^{(K)}}
\newcommand{\MyGammaMISO}{\Gamma^{(M_t)}_{(M_r)}}
\newcommand{\MyGammaSIMO}{\Gamma^{(M_r)}_{(M_t)}}

%%%%%%%%%%%%%%%%%%%%%%%%%%%%%%%%%%%%%%%%%%%%%%%%%
%%%%%%%%%%%%%%%%%%%%%%%%%%%%%%%%%%%%%%%%%%%%%%%%%
%%% TITLE %%%%%%%%%%%%%%%%%%%%%%%%%%%%%%%%%%%%%%%
%%%%%%%%%%%%%%%%%%%%%%%%%%%%%%%%%%%%%%%%%%%%%%%%%
%%%%%%%%%%%%%%%%%%%%%%%%%%%%%%%%%%%%%%%%%%%%%%%%%
\title{Characterization of Effective Capacity in Antenna Selection MIMO Systems}
\author{\IEEEauthorblockN{Mohammad~Lari\IEEEauthorrefmark{1},} \textit{Student Member, IEEE}, \and
        \IEEEauthorblockN{Abbas~Mohammadi\IEEEauthorrefmark{1},} \textit{Member, IEEE}, \and
        \IEEEauthorblockN{Abdolali~Abdipour\IEEEauthorrefmark{1},} \textit{Senior Member, IEEE}, \and
        \IEEEauthorblockN{Inkyu~Lee\IEEEauthorrefmark{2}} \textit{Senior Member, IEEE}, \\
        \IEEEauthorblockA{\IEEEauthorrefmark{1}Microwave, Millimeter Wave and Wireless Communications Research Lab., Electrical Engineering Department, Amirkabir University of Technology, Tehran, Iran}\\
        \IEEEauthorblockA{\IEEEauthorrefmark{2}School of Electrical Eng., Korea University, Seoul, S. Korea}}
\maketitle
%%%%%%%%%%%%%%%%%%%%%%%%%%%%%%%%%%%%%%%%%%%%%%%%%
%%%%%%%%%%%%%%%%%%%%%%%%%%%%%%%%%%%%%%%%%%%%%%%%%
%%% ABSTRACT %%%%%%%%%%%%%%%%%%%%%%%%%%%%%%%%%%%%
%%%%%%%%%%%%%%%%%%%%%%%%%%%%%%%%%%%%%%%%%%%%%%%%%
%%%%%%%%%%%%%%%%%%%%%%%%%%%%%%%%%%%%%%%%%%%%%%%%%
\begin{abstract}
In this Paper, the effective capacity of a multiple-input multiple-output (MIMO) system in two different cases with receive antenna selection (RAS) and transmit antenna selection (TAS) schemes is investigated. A closed-form solution for the maximum constant arrival rate of this network with statistical delay quality of service (QoS) constraint is extracted in the quasi-static fading channel. This study is conducted in two different cases. When channel state information (CSI) is not available at the MIMO transmitter, implementation of TAS is difficult. Therefore, RAS scheme is employed and one antenna with the maximum instantaneous signal to noise ratio (SNR) is chosen at the receiver. On the other hand, when CSI is available at the transmitter, TAS scheme is executed. In this case one antenna is selected at the transmitter. Moreover, an optimal power-control policy is applied to the selected antenna and the effective capacity of the MIMO system is derived. Finally, this optimal power adaptation and the effective capacity are investigated in two asymptotic cases with the loose and strict QoS requirements. In particular, we show that in the TAS scheme with the loose QoS restriction, the effective capacity converges to the ergodic capacity. Then, an exact closed-form solution is obtained for the ergodic capacity of the channel here.
\end{abstract}
%%%%%%%%%%%%%%%%%%%%%%%%%%%%%%%%%%%%%%%%%%%%%%%%%
%%%%%%%%%%%%%%%%%%%%%%%%%%%%%%%%%%%%%%%%%%%%%%%%%
%%% KEYWORDS %%%%%%%%%%%%%%%%%%%%%%%%%%%%%%%%%%%%
%%%%%%%%%%%%%%%%%%%%%%%%%%%%%%%%%%%%%%%%%%%%%%%%%
%%%%%%%%%%%%%%%%%%%%%%%%%%%%%%%%%%%%%%%%%%%%%%%%%
%%%%%%%%%%%%%%%%%%%%%%%%%%%%%%%%%%%%%%%%%%%%%%%%%
%%%%%%%%%%%%%%%%%%%%%%%%%%%%%%%%%%%%%%%%%%%%%%%%%
%%% SECTION 1 %%%%%%%%%%%%%%%%%%%%%%%%%%%%%%%%%%%
%%%%%%%%%%%%%%%%%%%%%%%%%%%%%%%%%%%%%%%%%%%%%%%%%
%%%%%%%%%%%%%%%%%%%%%%%%%%%%%%%%%%%%%%%%%%%%%%%%%
\section{Introduction}\label{sec:1}
With the increasing demand for high data-rate services in emerging communication systems such as third and fourth generation (3G and 4G) wireless networks, high-speed and high-quality data communications have become extremely necessary. Wireless channel is the major challenge in providing high-speed and high-quality data communications. Multiple-input multiple-output (MIMO) technology is an attractive solution to overcome these limitations. The multiple-antenna front-end architecture design, traditionally results in greater complexity and higher hardware costs in the radio frequency (RF) section \cite{LariBassam}-\cite{Mohammadi}. The complexity and cost generally increase with the increasing number of antennas. One simplifying and cost-reducing solution may be the utilization of a single RF front-end, where a single RF path is used instead of multiple parallel RF chains \cite{LariBassam}-\cite{Mohammadi}. In addition, antenna selection (AS) technique is a current method for dealing with this issue \cite{Bae}-\cite{Lari}. By using this approach, some of the available antennas at the transmitter and/or receiver are selected. Then, MIMO systems can use fewer RF chains than a number of transmit and/or receive antennas, which results in lower complexity and cost. AS can be implemented at the transmitter and/or receiver. In these schemes, an antenna can be selected to maximize the channel capacity \cite{Molisch} or to maximize the signal to noise ratio (SNR) \cite{Duman}. Likewise, cross-layer based AS approaches have been proposed recently \cite{AbuSaleh}-\cite{Lari}.

Multimedia applications such as video conferencing, require low end-to-end delay \cite{Pau}. For these type of applications delay becomes a critically important parameter for quality-of-service (QoS) provisioning. Once a delay requirement is violated, the corresponding data packet is discarded. In an appropriate channel condition, transmission with high data rate can be achieved through the wireless environment. In contrast, in a severe condition, the data rate decreases and probably becomes zero. Due to this, the statistical delay boundaries capable of characterizing the QoS requirements is employed for the delay-sensitive applications. For this purpose, effective capacity is defined as a maximum constant arrival rate that a wireless channel can support in order to guarantee the delay QoS requirements \cite{Negi}. In contrast to Shannon capacity without any restrictions on complexity and delay, the effective capacity ensures the maximum probabilistic delay for the incoming user traffic in the network \cite{Negi}.

For the wireless communication networks, the theory of effective capacity is proposed in \cite{Negi} and a closed-form solution is extracted for the correlated Rayleigh channel in the low-SNR regime. This new concept is completely discussed with a different viewpoint in \cite{Soret} for a generic source over correlated Rayleigh channels. The authors assume Gaussian distribution for the accumulated service rate of the channel which leads to the new result for the effective capacity over correlated Rayleigh channels. Following these fundamental research, the theory of effective capacity has been conducted in various systems and problems \cite{ZhangSISO}-\cite{Aboutorab}. The system throughput in a single-input single-output (SISO) channel subject to a given statistical delay QoS constraint is studied in \cite{ZhangSISO}. Then, an optimal power and rate adaptation policy which maximize the system throughput is derived. The discussion is also extended to a more practical scenario with variable-power adaptive modulation over both block fading and Markov correlated fading channels. In addition, a similar problem is considered in a MIMO system in \cite{ZhangMIMO}. In \cite{Femenias}, a novel framework based on the effective capacity theory for the cross-layer analysis and design of wireless networks combining adaptive modulation and coding (AMC) at the physical layer with an automatic repeat request (ARQ) protocol at the data-link layer is proposed. Moreover, a delay constrained performance of a cognitive radio and cognitive radio relay channels is evaluated in \cite{Aissa1} and \cite{Aissa2} respectively. In these papers the authors derive an optimal rate and power adaptation policy for the secondary user, which maximize the effective capacity subject to satisfying interference-power limitations on the primary user. An effective capacity based call admission control (CAC) with adaptive modulation technique to manage the self-similar traffic in a wireless IP-based network is also introduced in \cite{Aboutorab}.

MIMO systems involving AS technique at their transmitter or receiver, reveal significant advantages in practical communications. Hence, in this paper we study the performance of a MIMO system with the receive antenna selection (RAS) or transmit antenna selection (TAS) scheme for the delay sensitive user traffics. To the best of our knowledge, there is no prior work in this subject. Here, we assume an uncorrelated quasi-static Rayleigh fading channel in two different cases with the RAS and TAS schemes. In the RAS case, since there is no channel state information (CSI) available at the transmitter, AS at the transmitter is not possible. Thus, one antenna with the maximum instantaneous SNR is selected at the receiver. On the other hand, when the CSI is available at the transmitter, one antenna is chosen there. We select an antenna which maximizes the corresponding instantaneous SNR at the receiver. In order to improve the performance, an optimal power adaptation policy is considered for the selected antenna. In both cases with the RAS and TAS, we derive a closed-form expression for the effective capacity of the system. Finally, the optimal power and the effective capacity of the MIMO system with TAS scheme are analyzed in two asymptotic cases with the loose and high QoS constraints. In the former case with the loose QoS requirement, we indicate that the optimal power converges to a constant value with the high SNR assumption. Therefore the effective capacity with and without power adaptation are nearly the same. Consequently, the power adaptation policy does not have an advantage hare. We also show that the effective capacity approaches to the ergodic capacity and an exact closed-form expression for the ergodic capacity is derived here. The ergodic capacity can be viewed as an average of the Shannon capacity. On the other hand in the second case when high QoS is required, we observe that the TAS scheme can keep the effective capacity near the ergodic capacity too. This result is achieved while in a usual circumstance the effective capacity decreases when the QoS requirement increases. Therefore this can be so attractive for applications with the high QoS demands.

The rest of this paper is organized as follows: First, the system model is introduced in Section \ref{sec:2}. Section \ref{sec:3} provides a brief introduction to the effective capacity concept. The effective capacity of a MIMO system with both RAS and TAS schemes are explained in Section \ref{sec:4}. Then, the analysis of an optimal power control policy and the effective capacity in two asymptotic cases are also proposed here. Finally, the simulation results are presented in Section \ref{sec:5}, and Section \ref{sec:6} concludes the paper.
%%%%%%%%%%%%%%%%%%%%%%%%%%%%%%%%%%%%%%%%%%%%%%%%%
%%%%%%%%%%%%%%%%%%%%%%%%%%%%%%%%%%%%%%%%%%%%%%%%%
%%% SECTION 2 %%%%%%%%%%%%%%%%%%%%%%%%%%%%%%%%%%%
%%%%%%%%%%%%%%%%%%%%%%%%%%%%%%%%%%%%%%%%%%%%%%%%%
%%%%%%%%%%%%%%%%%%%%%%%%%%%%%%%%%%%%%%%%%%%%%%%%%
\section{System Model}\label{sec:2}
\begin{figure}
\begin{center}
\includegraphics[width=\linewidth]{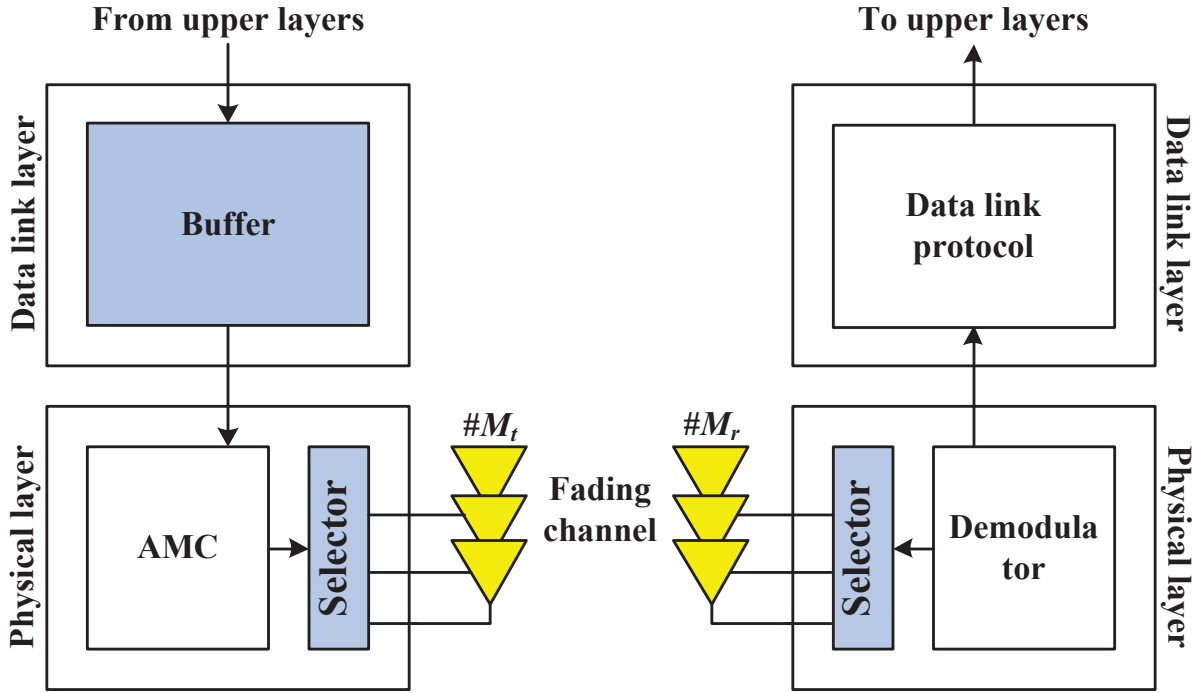}
\end{center}
\caption{System model.}
\label{fig:1}
\end{figure}

The system model is illustrated in Fig. \ref{fig:1}. First, information data from upper layers arrive to the data link layer where they are stored as an equal-length packet in the buffer. The packets are split up into bit streams and delivered to the physical layer. In the physical layer the bit streams are modulated with AMC scheme and make a frame. Then, with the RAS technique, the frames are transmitted from $M_t$ transmit antennas and received by the selected antenna at the receiver. However, with the TAS scheme, one antenna is chosen and transmits the frames. In such a case, all $M_r$ receive antennas capture the transmitted frames. At the physical layer in the receiver, the frames are demodulated and passed to the data link layer. Then, at the data link layer some error detection algorithms such as ARQ protocol are executed which is not considered in this paper. After that, data link layer sends the transmitted information data to upper layers. The detailed explanations of physical and data link layers are as follows.
%%%%%%%%%%%%%%%%%%%%%%%%%%%%%%%%%%%%%%%%%%%%%%%%%
%%%%%%%%%%%%%%%%%%%%%%%%%%%%%%%%%%%%%%%%%%%%%%%%%
%%% FIGURE 1 %%%%%%%%%%%%%%%%%%%%%%%%%%%%%%%%%%%%
%%%%%%%%%%%%%%%%%%%%%%%%%%%%%%%%%%%%%%%%%%%%%%%%%
%%%%%%%%%%%%%%%%%%%%%%%%%%%%%%%%%%%%%%%%%%%%%%%%%

The physical layer includes $M_t$ transmit and $M_r$ receive antennas in a MIMO channel. The link between each transmit and receive antenna is an uncorrelated quasi-static Rayleigh fading channel. We further assume that an ideal AMC scheme is implemented at the transmitter. Therefore, data can leave the transmitter with the instantaneous channel capacity rate. In the physical layer we adopt two different scenarios for the selection; RAS or TAS. In the RAS technique, we select one antenna with the maximum instantaneous SNR at the receiver. In this case the available power is identically distributed over the $M_t$ transmit antennas. In contrast in the TAS scheme, we select one antenna at the transmitter which maximizes the corresponding instantaneous SNR at the receiver. For further improvement, an optimal power can be used for the selected antenna.

At the data link layer of the transmitter, a simple first-input first-output (FIFO) buffer is assumed \cite{Jafari}. The source process of the network has a constant rate. Therefore, the buffer fills with a constant arrival rate and it is served with the instantaneous MIMO channel capacity rate. Since the channel capacity is time-varying, the service rate of the buffer is not constant. Hence, each packet needs to stay at the buffer for a while before transmission and this waiting time depends on the channel state. We also assume that enough packets are available in the buffer for transmission.

In the quasi-static fading, MIMO channel coefficients are constant during a frame time duration $T$. In this case the instantaneous SNR is defined as
%%%%%%%%%%%%%%%%%%%%%%%%%%%%%%%%%%%%%%%%%%%%%%%%%
%%%%%%%%%%%%%%%%%%%%%%%%%%%%%%%%%%%%%%%%%%%%%%%%%
%%% EQUATION 1 %%%%%%%%%%%%%%%%%%%%%%%%%%%%%%%%%%
%%%%%%%%%%%%%%%%%%%%%%%%%%%%%%%%%%%%%%%%%%%%%%%%%
%%%%%%%%%%%%%%%%%%%%%%%%%%%%%%%%%%%%%%%%%%%%%%%%%
\begin{equation}\label{eq:GammaK}
\MyGammaK=\Mygammazero\sum_{k=1}^{K}|h_k|^2,
\end{equation}
where in the RAS case, $K=M_t$ and $h_k$ denotes the channel coefficient between the $k$th transmit and a receive antenna. However, in the TAS case, $K=M_r$ and $h_k$ denotes the channel coefficient between a transmit and the $k$th receive antenna. In addition, $\Mygammazero=P_0/(N_0B)$ represents the average SNR; $P_0$ is the mean transmitted power; $B$ is the total spectral bandwidth of the system and $N_0/2$ stands for the power density per dimension of the additive white Gaussian noise (AWGN). The probability density function (PDF) and cumulative distribution function (CDF) of $\MyGammaK$ can be written as:
%%%%%%%%%%%%%%%%%%%%%%%%%%%%%%%%%%%%%%%%%%%%%%%%%
%%%%%%%%%%%%%%%%%%%%%%%%%%%%%%%%%%%%%%%%%%%%%%%%%
%%% EQUATION 2 %%%%%%%%%%%%%%%%%%%%%%%%%%%%%%%%%%
%%%%%%%%%%%%%%%%%%%%%%%%%%%%%%%%%%%%%%%%%%%%%%%%%
%%%%%%%%%%%%%%%%%%%%%%%%%%%%%%%%%%%%%%%%%%%%%%%%%
\begin{equation}\label{eq:pdf}
f_{\MyGammaK}(x)=\frac{1}{\Mygammazero^K\Gamma(K)}x^{K-1}e^{-x/\Mygammazero},~~~x\geq0
\end{equation}
%%%%%%%%%%%%%%%%%%%%%%%%%%%%%%%%%%%%%%%%%%%%%%%%%
%%%%%%%%%%%%%%%%%%%%%%%%%%%%%%%%%%%%%%%%%%%%%%%%%
%%% EQUATION 3 %%%%%%%%%%%%%%%%%%%%%%%%%%%%%%%%%%
%%%%%%%%%%%%%%%%%%%%%%%%%%%%%%%%%%%%%%%%%%%%%%%%%
%%%%%%%%%%%%%%%%%%%%%%%%%%%%%%%%%%%%%%%%%%%%%%%%%
\begin{equation}\label{eq:cdf}
F_{\MyGammaK}(x)=\frac{1}{\Gamma(K)}\gamma(K,x/\Mygammazero),~~~x\geq0,
\end{equation}
where $\Gamma(.)$ and $\gamma(.,.)$ denote the Gamma and the lower incomplete Gamma function, respectively.
In AS problems, we usually deal with the ordered random variables. Here $\{\MyGammaK_1,\MyGammaK_2,...,\MyGammaK_L\}$ denotes $L$ random variables defined in (\ref{eq:GammaK}), and $\{\MyGammaK_{(1)},\MyGammaK_{(2)},...,\MyGammaK_{(L)}\}$ represents ordered random variables where $\{\MyGammaK_{(1)}\leq\MyGammaK_{(2)}\leq...\leq\MyGammaK_{(L)}\}$. In the RAS case where the antenna with the maximum instantaneous SNR is selected, $L=M_r$ and similarly in the TAS case $L=M_t$. The PDF of $\MyGammaK_{(l)}$ and $1\leq l\leq L$ is given by \cite{David}
%%%%%%%%%%%%%%%%%%%%%%%%%%%%%%%%%%%%%%%%%%%%%%%%%
%%%%%%%%%%%%%%%%%%%%%%%%%%%%%%%%%%%%%%%%%%%%%%%%%
%%% EQUATION 4 %%%%%%%%%%%%%%%%%%%%%%%%%%%%%%%%%%
%%%%%%%%%%%%%%%%%%%%%%%%%%%%%%%%%%%%%%%%%%%%%%%%%
%%%%%%%%%%%%%%%%%%%%%%%%%%%%%%%%%%%%%%%%%%%%%%%%%
\begin{equation}\label{eq:generalOSpdf}
f_{(l)}(x)=\frac{L!}{(l-1)!(L-l)!}\left(F(x)\right)^{l-1}\left(1-F(x)\right)^{L-l}f(x)
\end{equation}
where for ease of notation, $f_{\MyGammaK}(x)$ and $F_{\MyGammaK}(x)$ are replaced by $f(x)$ and $F(x)$ in (\ref{eq:generalOSpdf}) and the subsequent expressions. For the maximum instantaneous SNR we also have \cite{David}
%%%%%%%%%%%%%%%%%%%%%%%%%%%%%%%%%%%%%%%%%%%%%%%%%
%%%%%%%%%%%%%%%%%%%%%%%%%%%%%%%%%%%%%%%%%%%%%%%%%
%%% EQUATION 5 %%%%%%%%%%%%%%%%%%%%%%%%%%%%%%%%%%
%%%%%%%%%%%%%%%%%%%%%%%%%%%%%%%%%%%%%%%%%%%%%%%%%
%%%%%%%%%%%%%%%%%%%%%%%%%%%%%%%%%%%%%%%%%%%%%%%%%
\begin{equation}\label{eq:maxOSpdf}
f_{(L)}(x)=L\left(F(x)\right)^{L-1}f(x).
\end{equation}
This PDF will be employed later to derive the effective capacity.
%%%%%%%%%%%%%%%%%%%%%%%%%%%%%%%%%%%%%%%%%%%%%%%%%
%%%%%%%%%%%%%%%%%%%%%%%%%%%%%%%%%%%%%%%%%%%%%%%%%
%%% SECTION 3 %%%%%%%%%%%%%%%%%%%%%%%%%%%%%%%%%%%
%%%%%%%%%%%%%%%%%%%%%%%%%%%%%%%%%%%%%%%%%%%%%%%%%
%%%%%%%%%%%%%%%%%%%%%%%%%%%%%%%%%%%%%%%%%%%%%%%%%
\section{The Effective Capacity}\label{sec:3}
Despite the time-varying nature of wireless channels, network service providers must guarantee a specified QoS to satisfy their customers who have real-time multimedia traffics. In \cite{Negi}, Wu and Negi defined the effective capacity as the maximum constant arrival rate that a given service process can support in order to guarantee a QoS requirement specified by the QoS exponent $\theta$. For a dynamic queuing system with stationary and ergodic arrival and service processes, under sufficient conditions, the queue length process $Q(t)$ converges in distribution to a random variable $Q(\infty)$ such that \cite{Negi}
%%%%%%%%%%%%%%%%%%%%%%%%%%%%%%%%%%%%%%%%%%%%%%%%%
%%%%%%%%%%%%%%%%%%%%%%%%%%%%%%%%%%%%%%%%%%%%%%%%%
%%% EQUATION 6 %%%%%%%%%%%%%%%%%%%%%%%%%%%%%%%%%%
%%%%%%%%%%%%%%%%%%%%%%%%%%%%%%%%%%%%%%%%%%%%%%%%%
%%%%%%%%%%%%%%%%%%%%%%%%%%%%%%%%%%%%%%%%%%%%%%%%%
\begin{equation}
-\lim_{q\to \infty}\frac{\ln\left(\textrm{Pr}\{Q(\infty)>q\}\right)}{q}=\theta,
\end{equation}
where $q$ determines a certain threshold, and therefore, we have the following approximation:
%%%%%%%%%%%%%%%%%%%%%%%%%%%%%%%%%%%%%%%%%%%%%%%%%
%%%%%%%%%%%%%%%%%%%%%%%%%%%%%%%%%%%%%%%%%%%%%%%%%
%%% EQUATION 7 %%%%%%%%%%%%%%%%%%%%%%%%%%%%%%%%%%
%%%%%%%%%%%%%%%%%%%%%%%%%%%%%%%%%%%%%%%%%%%%%%%%%
%%%%%%%%%%%%%%%%%%%%%%%%%%%%%%%%%%%%%%%%%%%%%%%%%
\begin{equation}
\textrm{Pr}\{Q(\infty)>q\}\approx e^{-\theta q}
\end{equation}
for a large $q$. For a small $q$ the following approximation is shown to be more accurate \cite{Negi}:
%%%%%%%%%%%%%%%%%%%%%%%%%%%%%%%%%%%%%%%%%%%%%%%%%
%%%%%%%%%%%%%%%%%%%%%%%%%%%%%%%%%%%%%%%%%%%%%%%%%
%%% EQUATION 8 %%%%%%%%%%%%%%%%%%%%%%%%%%%%%%%%%%
%%%%%%%%%%%%%%%%%%%%%%%%%%%%%%%%%%%%%%%%%%%%%%%%%
%%%%%%%%%%%%%%%%%%%%%%%%%%%%%%%%%%%%%%%%%%%%%%%%%
\begin{equation}\label{eq:a}
\textrm{Pr}\{Q(\infty)>q\}\approx \varepsilon e^{-\theta q},
\end{equation}
where $\varepsilon$ represents the non-empty buffer probability. In addition, for the delay experienced by a packet as the main QoS metric, we have a similar probability function as \cite{Negi}
%%%%%%%%%%%%%%%%%%%%%%%%%%%%%%%%%%%%%%%%%%%%%%%%%
%%%%%%%%%%%%%%%%%%%%%%%%%%%%%%%%%%%%%%%%%%%%%%%%%
%%% EQUATION 9 %%%%%%%%%%%%%%%%%%%%%%%%%%%%%%%%%%
%%%%%%%%%%%%%%%%%%%%%%%%%%%%%%%%%%%%%%%%%%%%%%%%%
%%%%%%%%%%%%%%%%%%%%%%%%%%%%%%%%%%%%%%%%%%%%%%%%%
\begin{equation}\label{eq:b}
\textrm{Pr}\{D>d\}\approx\varepsilon e^{-\theta \delta d},
\end{equation}
where $D$ indicates the tolerated delay, $d$ is a delay-bound, and $\delta$ is jointly determined by both arrival and service processes. The statistical delay constraint in (\ref{eq:b}) represents the QoS which has to be guaranteed for the delay sensitive traffic sources. It is apparent that the QoS exponent $\theta$ has an important role here. Larger $\theta$ corresponds to more strict QoS constraint, while smaller $\theta$ implies looser QoS requirements.

Effective capacity provides the maximum constant arrival rate that can be supported by the time-varying wireless channel under the statistical delay constraint (\ref{eq:b}). Since the average arrival rate is equal to the average departure rate when the buffer is in a steady-state \cite{Chang}, the effective capacity can be viewed as the maximum throughput in the presence of such a constraint. The effective capacity is defined in \cite[eq. 12]{Negi} and \cite[eq. 6]{Soret} in the correlated channels. However in the uncorrelated case it reduces to
%%%%%%%%%%%%%%%%%%%%%%%%%%%%%%%%%%%%%%%%%%%%%%%%%
%%%%%%%%%%%%%%%%%%%%%%%%%%%%%%%%%%%%%%%%%%%%%%%%%
%%% EQUATION 10 %%%%%%%%%%%%%%%%%%%%%%%%%%%%%%%%%
%%%%%%%%%%%%%%%%%%%%%%%%%%%%%%%%%%%%%%%%%%%%%%%%%
%%%%%%%%%%%%%%%%%%%%%%%%%%%%%%%%%%%%%%%%%%%%%%%%%
\begin{equation}\label{eq:generalEC}
E_C(\theta)=-\frac{1}{\theta}\ln\left(\mathbb{E}\left\{e^{-\theta R}\right\}\right)
\end{equation}
where $R$ is the time-varying rate of the channel and $\mathbb{E}\{.\}$ denotes the expectation. For a specific application with a given statistical delay requirement, the QoS exponent $\theta$ can be determined from (\ref{eq:b}). Then, the maximum constant arrival rate of the sources that a wireless channel can support in order to guarantee the given QoS, is determined from (\ref{eq:generalEC}). $R$ and the effective capacity $E_C(\theta)$ are further discussed for two different AS schemes in the next section.
%%%%%%%%%%%%%%%%%%%%%%%%%%%%%%%%%%%%%%%%%%%%%%%%%
%%%%%%%%%%%%%%%%%%%%%%%%%%%%%%%%%%%%%%%%%%%%%%%%%
%%% SECTION 4 %%%%%%%%%%%%%%%%%%%%%%%%%%%%%%%%%%%
%%%%%%%%%%%%%%%%%%%%%%%%%%%%%%%%%%%%%%%%%%%%%%%%%
%%%%%%%%%%%%%%%%%%%%%%%%%%%%%%%%%%%%%%%%%%%%%%%%%
\section{Effective Capacity with Antenna Selection}\label{sec:4}
In the MIMO system with AS technique, we select one antenna at the receiver or one antenna at the transmitter in the RAS or TAS scheme respectively. Single antenna selection is the simplest method for implementing this technique \cite{Coskunand}. In order to eliminate practical issues, such as mutual coupling, spatial correlation of antennas and inaccurate time-synchronization of the antennas, single antenna selection is extensively considered in the literatures \cite{Catreux}-\cite{Blum}. In a MIMO system with the single AS, we deal with a multiple-input single-output (MISO) system in the RAS case, and a single-input multiple-output (SIMO) system in the TAS case. The effective capacity of these systems are studied in the following sections more precisely.
%%%%%%%%%%%%%%%%%%%%%%%%%%%%%%%%%%%%%%%%%%%%%%%%%
%%%%%%%%%%%%%%%%%%%%%%%%%%%%%%%%%%%%%%%%%%%%%%%%%
%%% SUBSECTION A %%%%%%%%%%%%%%%%%%%%%%%%%%%%%%%%
%%%%%%%%%%%%%%%%%%%%%%%%%%%%%%%%%%%%%%%%%%%%%%%%%
%%%%%%%%%%%%%%%%%%%%%%%%%%%%%%%%%%%%%%%%%%%%%%%%%
\subsection{Receive Antenna Selection}\label{subsec:A}
In the RAS case with $M_t$ transmit and one selected receive antennas, $\{\Gamma^{(M_t)}_1,\Gamma^{(M_t)}_2,...,\Gamma^{(M_t)}_{M_r}\}$ represent the instantaneous SNRs of $M_r$ available receive antennas and $\{\Gamma^{(M_t)}_{(1)},\Gamma^{(M_t)}_{(2)},...,\Gamma^{(M_t)}_{(M_r)}\}$ show the ordered random variables where $\{\Gamma^{(M_t)}_{(1)}\leq\Gamma^{(M_t)}_{(2)}\leq...\leq\Gamma^{(M_t)}_{(M_r)}\}$. Here we choose one antenna with the maximum instantaneous SNR. This means that the corresponding antenna to $\Gamma^{(M_t)}_{(M_r)}$ will be used for transmission. This antenna has a maximum capacity too. Now, the service rate in a MISO system can be denoted by \cite{Paulraj}
%%%%%%%%%%%%%%%%%%%%%%%%%%%%%%%%%%%%%%%%%%%%%%%%%
%%%%%%%%%%%%%%%%%%%%%%%%%%%%%%%%%%%%%%%%%%%%%%%%%
%%% EQUATION 11 %%%%%%%%%%%%%%%%%%%%%%%%%%%%%%%%%
%%%%%%%%%%%%%%%%%%%%%%%%%%%%%%%%%%%%%%%%%%%%%%%%%
%%%%%%%%%%%%%%%%%%%%%%%%%%%%%%%%%%%%%%%%%%%%%%%%%
\begin{equation}\label{eq:MISOcapacity}
R=BT\log_2\left(1+\frac{1}{M_t}\MyGammaMISO\right),
\end{equation}
where $B$ is the total spectral bandwidth and $T$ is the frame duration. By inserting (\ref{eq:MISOcapacity}) into (\ref{eq:generalEC}), the effective capacity can be found according to
%%%%%%%%%%%%%%%%%%%%%%%%%%%%%%%%%%%%%%%%%%%%%%%%%
%%%%%%%%%%%%%%%%%%%%%%%%%%%%%%%%%%%%%%%%%%%%%%%%%
%%% EQUATION 12 %%%%%%%%%%%%%%%%%%%%%%%%%%%%%%%%%
%%%%%%%%%%%%%%%%%%%%%%%%%%%%%%%%%%%%%%%%%%%%%%%%%
%%%%%%%%%%%%%%%%%%%%%%%%%%%%%%%%%%%%%%%%%%%%%%%%%
\begin{equation}\label{eq:generalECMISO}
E_C(\theta)=-\frac{1}{\theta}\ln\left(\mathbb{E}\left\{\left(1+\frac{1}{M_t}\MyGammaMISO\right)^{-\Mythetat}\right\}\right),
\end{equation}
where $\Mythetat=\theta BT/\ln(2)$.

The PDF of $\MyGammaMISO$ is introduced in (\ref{eq:maxOSpdf}). However here for our purpose, we use the binomial and multinomial theorems and extract a modified representation for the PDF. In the RAS case $L=M_r$ and $K=M_t$ is replaced in (\ref{eq:maxOSpdf}) for the following operations. When the first argument of the lower incomplete Gamma function is an integer, this function can be written as a finite summation. Therefore using \cite[eq. 8.352-1]{Ryzhik}, we can rewrite (\ref{eq:cdf}) as
%%%%%%%%%%%%%%%%%%%%%%%%%%%%%%%%%%%%%%%%%%%%%%%%%
%%%%%%%%%%%%%%%%%%%%%%%%%%%%%%%%%%%%%%%%%%%%%%%%%
%%% EQUATION 13 %%%%%%%%%%%%%%%%%%%%%%%%%%%%%%%%%
%%%%%%%%%%%%%%%%%%%%%%%%%%%%%%%%%%%%%%%%%%%%%%%%%
%%%%%%%%%%%%%%%%%%%%%%%%%%%%%%%%%%%%%%%%%%%%%%%%%
\begin{equation}\label{eq:simplifiedcdf1}
F(x)=1-e^{-x/\Mygammazero}\sum_{k=0}^{M_t-1}\frac{x^k}{\Mygammazero^kk!}.
\end{equation}
Then, through using the binomial theorem we obtain
%%%%%%%%%%%%%%%%%%%%%%%%%%%%%%%%%%%%%%%%%%%%%%%%%
%%%%%%%%%%%%%%%%%%%%%%%%%%%%%%%%%%%%%%%%%%%%%%%%%
%%% EQUATION 14 %%%%%%%%%%%%%%%%%%%%%%%%%%%%%%%%%
%%%%%%%%%%%%%%%%%%%%%%%%%%%%%%%%%%%%%%%%%%%%%%%%%
%%%%%%%%%%%%%%%%%%%%%%%%%%%%%%%%%%%%%%%%%%%%%%%%%
\begin{align}\label{eq:simplidiedcdf2}
&\left(F(x)\right)^{M_r-1}=\nonumber\\
&\sum_{m=0}^{M_r-1}{M_r-1\choose m}(-1)^me^{-mx/\Mygammazero}\left(\sum_{k=0}^{M_t-1}\frac{x^k}{\Mygammazero^kk!}\right)^m.
\end{align}
By employing the multinomial theorem, (\ref{eq:simplidiedcdf2}) finally reduces to
%%%%%%%%%%%%%%%%%%%%%%%%%%%%%%%%%%%%%%%%%%%%%%%%%
%%%%%%%%%%%%%%%%%%%%%%%%%%%%%%%%%%%%%%%%%%%%%%%%%
%%% EQUATION 15 %%%%%%%%%%%%%%%%%%%%%%%%%%%%%%%%%
%%%%%%%%%%%%%%%%%%%%%%%%%%%%%%%%%%%%%%%%%%%%%%%%%
%%%%%%%%%%%%%%%%%%%%%%%%%%%%%%%%%%%%%%%%%%%%%%%%%
\begin{equation}\label{eq:simplidiedcdf3}
\left(F(x)\right)^{M_r-1}=\sum_{m=0}^{M_r-1}{M_r-1\choose m}(-1)^me^{-mx/\Mygammazero}\sum_{q=0}^{Q(m)}c_{q}^{(m)}x^q,
\end{equation}
where $Q(m)=m(M_t-1)$, and $c_q^{(m)}$ is the resultant coefficients from the multinomial expansion of $\left(\sum_{k=0}^{M_t-1}\frac{x^{k}}{\Mygammazero^{k}{k}!}\right)^{m}$. In our case, $c_q^{(m)}$ is directly calculated with discrete convolution. $\vec{\textbf{c}}^{(0)}=1$, and $\vec{\textbf{c}}^{(1)}$ can be defined as
%%%%%%%%%%%%%%%%%%%%%%%%%%%%%%%%%%%%%%%%%%%%%%%%%
%%%%%%%%%%%%%%%%%%%%%%%%%%%%%%%%%%%%%%%%%%%%%%%%%
%%% EQUATION 16 %%%%%%%%%%%%%%%%%%%%%%%%%%%%%%%%%
%%%%%%%%%%%%%%%%%%%%%%%%%%%%%%%%%%%%%%%%%%%%%%%%%
%%%%%%%%%%%%%%%%%%%%%%%%%%%%%%%%%%%%%%%%%%%%%%%%%
\begin{equation}\label{eq:c1}
\vec{\textbf{c}}^{(1)}=\left[
                       \begin{array}{ccccc}
                         1 & \frac{1}{\Mygammazero 1!} & \frac{1}{\Mygammazero^2 2!} & ... & \frac{1}{\Mygammazero^{M_t-1}(M_t-1)!} \\
                       \end{array}
                     \right]
\end{equation}
and therefore, $c_q^{(m)}$ is the $q$th element of the vector $\vec{\textbf{c}}^{(m)}$ where $\vec{\textbf{c}}^{(m)}=\vec{\textbf{c}}^{(m-1)}\otimes\vec{\textbf{c}}^{(1)}$, and $\otimes$ represents the  discrete convolution function. Finally, the PDF of $\MyGammaMISO$ is obtained as
%%%%%%%%%%%%%%%%%%%%%%%%%%%%%%%%%%%%%%%%%%%%%%%%%
%%%%%%%%%%%%%%%%%%%%%%%%%%%%%%%%%%%%%%%%%%%%%%%%%
%%% EQUATION 17 %%%%%%%%%%%%%%%%%%%%%%%%%%%%%%%%%
%%%%%%%%%%%%%%%%%%%%%%%%%%%%%%%%%%%%%%%%%%%%%%%%%
%%%%%%%%%%%%%%%%%%%%%%%%%%%%%%%%%%%%%%%%%%%%%%%%%
\begin{align}\label{eq:finalOSpdf}
&f_{(M_r)}(x)=\frac{M_r}{\Mygammazero^{M_t}\Gamma(M_t)}\times\nonumber\\
&\sum_{m=0}^{M_r-1}{M_r-1\choose m}(-1)^{m}\sum_{q=0}^{Q(m)}c_q^{(m)}x^{q+M_t-1}e^{-(m+1)x/\Mygammazero}.
\end{align}
Now, the expected value in (\ref{eq:generalECMISO}) is attained as
%%%%%%%%%%%%%%%%%%%%%%%%%%%%%%%%%%%%%%%%%%%%%%%%%
%%%%%%%%%%%%%%%%%%%%%%%%%%%%%%%%%%%%%%%%%%%%%%%%%
%%% EQUATION 18 %%%%%%%%%%%%%%%%%%%%%%%%%%%%%%%%%
%%%%%%%%%%%%%%%%%%%%%%%%%%%%%%%%%%%%%%%%%%%%%%%%%
%%%%%%%%%%%%%%%%%%%%%%%%%%%%%%%%%%%%%%%%%%%%%%%%%
\begin{align}\label{eq:expectationECMISO}
&\mathbb{E}\left\{\left(1+\frac{1}{M_t}\MyGammaMISO\right)^{-\Mythetat}\right\}=\int_{0}^{+\infty}\left(1+\frac{x}{M_t}\right)^{-\Mythetat}f_{(M_r)}(x)dx\nonumber\\
&=\frac{M_r}{\Mygammazero^{M_t}\Gamma(M_t)}\sum_{m=0}^{M_r-1}{M_r-1\choose m}(-1)^{m}\left[\sum_{q=0}^{Q(m)}c_q^{(m)}M_t^{M_t+q}\times\right.\nonumber\\
&\left.\Gamma(M_t+q)\psi\left(M_t+q,M_t+q-\Mythetat+1;\frac{(m+1)M_t}{\Mygammazero}\right)\vphantom{\sum_{q=0}^{Q(m)}}\right],
\end{align}
where $\psi(.,.;.)$ represents the confluent hypergeometric function \cite[eq. 9.210-2]{Ryzhik}. Applying the preceding expression to (\ref{eq:generalECMISO}) we get
%%%%%%%%%%%%%%%%%%%%%%%%%%%%%%%%%%%%%%%%%%%%%%%%%
%%%%%%%%%%%%%%%%%%%%%%%%%%%%%%%%%%%%%%%%%%%%%%%%%
%%% EQUATION 19 %%%%%%%%%%%%%%%%%%%%%%%%%%%%%%%%%
%%%%%%%%%%%%%%%%%%%%%%%%%%%%%%%%%%%%%%%%%%%%%%%%%
%%%%%%%%%%%%%%%%%%%%%%%%%%%%%%%%%%%%%%%%%%%%%%%%%
\begin{align}
&E_C(\theta)=-\frac{1}{\theta}\ln\left(\frac{M_r}{\Mygammazero^{M_t}\Gamma(M_t)}\sum_{m=0}^{M_r-1}{M_r-1\choose m}(-1)^{m}\times\right.\nonumber\\
&\left[\sum_{q=0}^{Q(m)}c_q^{(m)}M_t^{M_t+q}\Gamma(M_t+q)\times\right.\nonumber\\
&\left.\left.\psi\left(M_t+q,M_t+q-\frac{\theta BT}{\ln2}+1;\frac{(m+1)M_t}{\Mygammazero}\right)\vphantom{\sum_{q=0}^{Q(m)}}\right]\vphantom{\sum_{m=0}^{M_r-1}}\right).
\end{align}
Here we use
%%%%%%%%%%%%%%%%%%%%%%%%%%%%%%%%%%%%%%%%%%%%%%%%%
%%%%%%%%%%%%%%%%%%%%%%%%%%%%%%%%%%%%%%%%%%%%%%%%%
%%% EQUATION 20 %%%%%%%%%%%%%%%%%%%%%%%%%%%%%%%%%
%%%%%%%%%%%%%%%%%%%%%%%%%%%%%%%%%%%%%%%%%%%%%%%%%
%%%%%%%%%%%%%%%%%%%%%%%%%%%%%%%%%%%%%%%%%%%%%%%%%
\begin{equation}
\psi(a,b;z)=\frac{1}{\Gamma(a)}\int_{0}^{\infty}e^{-zt}t^{a-1}(1+t)^{b-a-1}dt
\end{equation}
as an integral representation of the confluent hypergeometric function \cite[eq, 9.211-4]{Ryzhik}.
%%%%%%%%%%%%%%%%%%%%%%%%%%%%%%%%%%%%%%%%%%%%%%%%%
%%%%%%%%%%%%%%%%%%%%%%%%%%%%%%%%%%%%%%%%%%%%%%%%%
%%% SUBSECTION B %%%%%%%%%%%%%%%%%%%%%%%%%%%%%%%%
%%%%%%%%%%%%%%%%%%%%%%%%%%%%%%%%%%%%%%%%%%%%%%%%%
%%%%%%%%%%%%%%%%%%%%%%%%%%%%%%%%%%%%%%%%%%%%%%%%%
\subsection{Transmit Antenna Selection}\label{subsec:B}
In a similar way, in the TAS case we assume one selected antenna at the transmitter and $M_r$ antennas at the receiver. $\{\Gamma^{(M_r)}_1,\Gamma^{(M_r)}_2,...,\Gamma^{(M_r)}_{M_t}\}$ represent the instantaneous SNRs at the receiver according to $M_t$ available transmit antennas and $\{\Gamma^{(M_r)}_{(1)},\Gamma^{(M_r)}_{(2)},...,\Gamma^{(M_r)}_{(M_t)}\}$ indicate the ordered random variables where $\{\Gamma^{(M_r)}_{(1)}\leq\Gamma^{(M_r)}_{(2)}\leq...\leq\Gamma^{(M_r)}_{(M_t)}\}$. Here, the corresponding antenna to $\Gamma^{(M_r)}_{(M_t)}$ is chosen. This antenna has a maximum capacity too. In addition, an optimal power adaptation is also assumed for the selected antenna. Now, the service rate in a SIMO system can be written as \cite{Paulraj}
%%%%%%%%%%%%%%%%%%%%%%%%%%%%%%%%%%%%%%%%%%%%%%%%%
%%%%%%%%%%%%%%%%%%%%%%%%%%%%%%%%%%%%%%%%%%%%%%%%%
%%% EQUATION 21 %%%%%%%%%%%%%%%%%%%%%%%%%%%%%%%%%
%%%%%%%%%%%%%%%%%%%%%%%%%%%%%%%%%%%%%%%%%%%%%%%%%
%%%%%%%%%%%%%%%%%%%%%%%%%%%%%%%%%%%%%%%%%%%%%%%%%
\begin{equation}\label{eq:SIMOcapacity}
R=BT\log_2\left(1+\mu\left[\MyGammaSIMO,\theta\right]\MyGammaSIMO\right),
\end{equation}
where $\mu\left[\MyGammaSIMO,\theta\right]$ is a coefficient which determines the optimal transmitted power over the time as $\mu\left[\MyGammaSIMO,\theta\right]P_0$. Note that $\mu\left[\MyGammaSIMO,\theta\right]$ is a function of the instantaneous SNR, $\MyGammaSIMO$, and also QoS exponent $\theta$. However, for ease of notation, we use this coefficient as $\mu$. For the constant average power, we must have $\mathbb{E}\{\mu\}=1$. The PDF of $\MyGammaSIMO$ is given in (\ref{eq:maxOSpdf}). By inserting $L=M_t$ and $K=M_r$ into (\ref{eq:maxOSpdf}) and using the binomial and mutinomial theorems once again, the modified representation for the PDF is achieved similar to (\ref{eq:finalOSpdf}) and used here for further analysis.

For the optimal value of $\mu$, the following optimization problem
%%%%%%%%%%%%%%%%%%%%%%%%%%%%%%%%%%%%%%%%%%%%%%%%%
%%%%%%%%%%%%%%%%%%%%%%%%%%%%%%%%%%%%%%%%%%%%%%%%%
%%% EQUATION 22 %%%%%%%%%%%%%%%%%%%%%%%%%%%%%%%%%
%%%%%%%%%%%%%%%%%%%%%%%%%%%%%%%%%%%%%%%%%%%%%%%%%
%%%%%%%%%%%%%%%%%%%%%%%%%%%%%%%%%%%%%%%%%%%%%%%%%
\begin{align}\label{eq:optproblem}
&\mu=\textrm{arg}\max_{\mu}\left(-\frac{1}{\theta}\ln \mathbb{E}\left\{\left(1+\mu\MyGammaSIMO\right)^{-\Mythetat}\right\}\right)\nonumber\\
&\textrm{s.t.}~~~\mathbb{E}\{\mu\}=1
\end{align}
has to be solved. Using the Lagrangian optimization method, the optimal solution is found as \cite{ZhangSISO}
%%%%%%%%%%%%%%%%%%%%%%%%%%%%%%%%%%%%%%%%%%%%%%%%%
%%%%%%%%%%%%%%%%%%%%%%%%%%%%%%%%%%%%%%%%%%%%%%%%%
%%% EQUATION 23 %%%%%%%%%%%%%%%%%%%%%%%%%%%%%%%%%
%%%%%%%%%%%%%%%%%%%%%%%%%%%%%%%%%%%%%%%%%%%%%%%%%
%%%%%%%%%%%%%%%%%%%%%%%%%%%%%%%%%%%%%%%%%%%%%%%%%
\begin{equation}\label{eq:mu}
\mu=
\begin{cases}
0&,\MyGammaSIMO<\MyGammazero\\
\left(\frac{1}{\MyGammazero}\right)^{\frac{1}{\Mythetat+1}}\left(\frac{1}{\MyGammaSIMO}\right)^{\frac{\Mythetat}{\Mythetat+1}}-\frac{1}{\MyGammaSIMO}&,\MyGammaSIMO\geq\MyGammazero
\end{cases}
\end{equation}
where $\MyGammazero=\lambda/\tilde{\theta}$ is a cutoff SNR threshold, which can be obtained from the average power constraint $\mathbb{E}\{\mu\}=1$ and $\lambda$ denotes Lagrangian coefficient. Fortunately, the mean value has a closed-form solution as (\ref{eq:expextationmu})
%%%%%%%%%%%%%%%%%%%%%%%%%%%%%%%%%%%%%%%%%%%%%%%%%
%%%%%%%%%%%%%%%%%%%%%%%%%%%%%%%%%%%%%%%%%%%%%%%%%
%%% EQUATION 24 %%%%%%%%%%%%%%%%%%%%%%%%%%%%%%%%%
%%%%%%%%%%%%%%%%%%%%%%%%%%%%%%%%%%%%%%%%%%%%%%%%%
%%%%%%%%%%%%%%%%%%%%%%%%%%%%%%%%%%%%%%%%%%%%%%%%%
\begin{figure*}
\begin{align}\label{eq:expextationmu}
&\mathbb{E}\{\mu\}=\int_{\MyGammazero}^{+\infty}\mu f_{(M_t)}(x)dx\nonumber\\
&=\frac{M_t}{\Mygammazero^{M_r}\Gamma(M_r)}\sum_{m=0}^{M_t-1}{M_t-1\choose m}(-1)^m\sum_{q=0}^{Q(m)}c_q^{(m)}\int_{\MyGammazero}^{+\infty}\left(\left(\frac{1}{\MyGammazero}\right)^{\frac{1}{\Mythetat+1}}\left(\frac{1}{x}\right)^{\frac{\Mythetat}{\Mythetat+1}}-\left(\frac{1}{x}\right)\right)x^{M_r+q-1}e^{-\frac{(m+1)}{\Mygammazero}x}dx\nonumber\\
&=\frac{M_t}{\Mygammazero^{M_r}\Gamma(M_r)}\sum_{m=0}^{M_t-1}{M_t-1\choose m}(-1)^m\sum_{q=0}^{Q(m)}c_q^{(m)}\left[\left(\frac{1}{\MyGammazero}\right)^{\frac{1}{\Mythetat+1}}\left(\frac{m+1}{\Mygammazero}\right)^{-\left(M_r+q-\frac{\Mythetat}{\Mythetat+1}\right)}\right.\times\nonumber\\
&\left.~~~\Gamma\left(M_r+q-\frac{\Mythetat}{\Mythetat+1},\frac{(m+1)\MyGammazero}{\Mygammazero}\right)-\left(\frac{m+1}{\Mygammazero}\right)^{-(M_r+q-1)}\Gamma\left(M_r+q-1,\frac{(m+1)\MyGammazero}{\Mygammazero}\right)\right]=1
\end{align}
\line(1,0){525}
\end{figure*}
where $\Gamma(.,.)$ represents the upper incomplete Gamma function. To derive (\ref{eq:expextationmu}) we used the following equality \cite[eq. 3.381-3]{Ryzhik}
%%%%%%%%%%%%%%%%%%%%%%%%%%%%%%%%%%%%%%%%%%%%%%%%%
%%%%%%%%%%%%%%%%%%%%%%%%%%%%%%%%%%%%%%%%%%%%%%%%%
%%% EQUATION 25 %%%%%%%%%%%%%%%%%%%%%%%%%%%%%%%%%
%%%%%%%%%%%%%%%%%%%%%%%%%%%%%%%%%%%%%%%%%%%%%%%%%
%%%%%%%%%%%%%%%%%%%%%%%%%%%%%%%%%%%%%%%%%%%%%%%%%
\begin{equation}
\int_{u}^{+\infty}x^{a-1}e^{-bx}dx=b^{-a}\Gamma(a,bu).
\end{equation}
In order to find the cutoff SNR threshold, $\MyGammazero$, we can use (\ref{eq:expextationmu}) where $\MyGammazero$ is calculated numerically. However, this numerical method is so simpler than the calculation of $\MyGammazero$ directly from the integral equation $\mathbb{E}\{\mu\}=1$.

Now, the expected value in (\ref{eq:optproblem}) can be expressed as (\ref{eq:expectationECSIMO})
%%%%%%%%%%%%%%%%%%%%%%%%%%%%%%%%%%%%%%%%%%%%%%%%%
%%%%%%%%%%%%%%%%%%%%%%%%%%%%%%%%%%%%%%%%%%%%%%%%%
%%% EQUATION 26 %%%%%%%%%%%%%%%%%%%%%%%%%%%%%%%%%
%%%%%%%%%%%%%%%%%%%%%%%%%%%%%%%%%%%%%%%%%%%%%%%%%
%%%%%%%%%%%%%%%%%%%%%%%%%%%%%%%%%%%%%%%%%%%%%%%%%
\begin{figure*}
\begin{align}\label{eq:expectationECSIMO}
&\mathbb{E}\left\{\left(1+\mu\MyGammaSIMO\right)^{-\Mythetat}\right\}=\int_{0}^{\MyGammazero}f_{(M_t)}(x)dx+\int_{\MyGammazero}^{+\infty}\left(\frac{x}{\MyGammazero}\right)^{-\frac{\Mythetat}{\Mythetat+1}}f_{(M_t)}(x)dx\nonumber\\
&=\frac{M_t}{\Mygammazero^{M_r}\Gamma(M_r)}\sum_{m=0}^{M_t-1}{M_t-1\choose m}(-1)^m\sum_{q=0}^{Q(m)}c_q^{(m)}\left[\left(\frac{m+1}{\Mygammazero}\right)^{-(M_r+q)}\gamma\left(M_r+q,\frac{(m+1)\MyGammazero}{\Mygammazero}\right)+\right.\nonumber\\
&~~~\left.\left(\frac{1}{\MyGammazero}\right)^{-\frac{\Mythetat}{\Mythetat+1}}\left(\frac{m+1}{\Mygammazero}\right)^{-\left(M_r+q-\frac{\Mythetat}{\Mythetat+1}\right)}\Gamma\left(M_r+q-\frac{\Mythetat}{\Mythetat+1},\frac{(m+1)\MyGammazero}{\Mygammazero}\right)\right]
\end{align}
\line(1,0){525}
\end{figure*}
and the effective capacity as
%%%%%%%%%%%%%%%%%%%%%%%%%%%%%%%%%%%%%%%%%%%%%%%%%
%%%%%%%%%%%%%%%%%%%%%%%%%%%%%%%%%%%%%%%%%%%%%%%%%
%%% EQUATION 27 %%%%%%%%%%%%%%%%%%%%%%%%%%%%%%%%%
%%%%%%%%%%%%%%%%%%%%%%%%%%%%%%%%%%%%%%%%%%%%%%%%%
%%%%%%%%%%%%%%%%%%%%%%%%%%%%%%%%%%%%%%%%%%%%%%%%%
\begin{equation}\label{eq:ECSIMO}
E_C(\theta)=-\frac{1}{\theta}\ln\left(\mathbb{E}\left\{\left(1+\mu\MyGammaSIMO\right)^{-\theta(BT/\ln2)}\right\}\right)
\end{equation}
where $Q(m)=m(M_r-1)$, and $c_q^{(m)}$ represents the multinomial coefficient. Note that, a closed-form solution for (\ref{eq:expectationECSIMO}) can be found by using \cite[eq. 3.381-3]{Ryzhik} and
%%%%%%%%%%%%%%%%%%%%%%%%%%%%%%%%%%%%%%%%%%%%%%%%%
%%%%%%%%%%%%%%%%%%%%%%%%%%%%%%%%%%%%%%%%%%%%%%%%%
%%% EQUATION 28 %%%%%%%%%%%%%%%%%%%%%%%%%%%%%%%%%
%%%%%%%%%%%%%%%%%%%%%%%%%%%%%%%%%%%%%%%%%%%%%%%%%
%%%%%%%%%%%%%%%%%%%%%%%%%%%%%%%%%%%%%%%%%%%%%%%%%
\begin{equation}
\int_{0}^{u}x^{a-1}e^{-bx}dx=b^{-a}\gamma(a,bu),
\end{equation}
for the lower incomplete Gamma function \cite[eq. 3.381-1]{Ryzhik}.

In a specific case when joint transmit and receive antenna selection is applied, it is sufficient to replace $M_t$ by $M_tM_r$ and $M_r$ by one in (\ref{eq:expextationmu}), (\ref{eq:expectationECSIMO}), and (\ref{eq:ECSIMO}), respectively.
%%%%%%%%%%%%%%%%%%%%%%%%%%%%%%%%%%%%%%%%%%%%%%%%%
%%%%%%%%%%%%%%%%%%%%%%%%%%%%%%%%%%%%%%%%%%%%%%%%%
%%% SUBSECTION C %%%%%%%%%%%%%%%%%%%%%%%%%%%%%%%%
%%%%%%%%%%%%%%%%%%%%%%%%%%%%%%%%%%%%%%%%%%%%%%%%%
%%%%%%%%%%%%%%%%%%%%%%%%%%%%%%%%%%%%%%%%%%%%%%%%%
\subsection{Asymptotic Analysis}\label{subsec:C}
Behavior of the optimal power coefficient $\mu$ and the effective capacity $E_C(\theta)$ in a MIMO system with the TAS scheme is discussed here. In the first case, when $\theta\to 0$, from (\ref{eq:mu}) we have
%%%%%%%%%%%%%%%%%%%%%%%%%%%%%%%%%%%%%%%%%%%%%%%%%
%%%%%%%%%%%%%%%%%%%%%%%%%%%%%%%%%%%%%%%%%%%%%%%%%
%%% EQUATION 29 %%%%%%%%%%%%%%%%%%%%%%%%%%%%%%%%%
%%%%%%%%%%%%%%%%%%%%%%%%%%%%%%%%%%%%%%%%%%%%%%%%%
%%%%%%%%%%%%%%%%%%%%%%%%%%%%%%%%%%%%%%%%%%%%%%%%%
\begin{equation}\label{eq:limthetazero}
\lim_{\theta\to 0}\mu=
\begin{cases}
0&,\MyGammaSIMO<\MyGammazero\\
\frac{1}{\MyGammazero}-\frac{1}{\MyGammaSIMO}&,\MyGammaSIMO\geq\MyGammazero
\end{cases}
\end{equation}
which depicts the water-filling formula similar to \cite[eq. 10]{ZhangSISO}. In addition when $\theta\to 0$ and $\Mygammazero\to\infty$, the optimal coefficient $\mu$ converges to a constant $\mu=1$. Therefore, the transmitted power $\mu P_0$ converges to a constant $P_0$ and the power adaptation policy does not have an advantage here. The proofs for the special cases of $M_t=1,~2$ with an arbitrary number of receive antennas, and $M_r=1,~2$ with an arbitrary number of transmit antennas, are provided in the Appendix \ref{appendix:1}. Simulation results indicate that this convergence is also true in a general case with arbitrary number of transmit and receive antennas and also for the moderate average SNRs.

In the time varying channel, the ergodic capacity defines the average rate which can be passed through this environment without any constraints on the QoS. Therefore, the ergodic capacity can be considered as an upper bound on the effective capacity. This significant subject is investigated extensively in literatures such as \cite{Nosratinia} with the asymptote of large number of transmit antennas and high SNR regime. Here when $\theta\to 0$, the effective capacity converges to the ergodic capacity $E_C^{(0)}$. In Appendix \ref{appendix:2} we first prove that the ergodic capacity can be achieved from the effective capacity when $\theta\to 0$ and then an exact closed-form solution for the ergodic capacity of a MIMO system with the TAS scheme and optimal power adaptation is extracted in (\ref{eq:50}). To derive this expression, we do not use any additional assumptions like \cite{Nosratinia}. Therefore, this expression is valid for all conditions.

In the second asymptotic case, we assume that $\theta\to\infty$. In the optimization problem (\ref{eq:optproblem}) we have $\MyGammazero=\lambda/\tilde{\theta}$. Therefore when $\theta\to\infty$ then $\MyGammazero\to 0$. Using (\ref{eq:mu}) and the mean power constraint $\mathbb{E}\{\mu\}=1$, we can write
%%%%%%%%%%%%%%%%%%%%%%%%%%%%%%%%%%%%%%%%%%%%%%%%%
%%%%%%%%%%%%%%%%%%%%%%%%%%%%%%%%%%%%%%%%%%%%%%%%%
%%% EQUATION 30 %%%%%%%%%%%%%%%%%%%%%%%%%%%%%%%%%
%%%%%%%%%%%%%%%%%%%%%%%%%%%%%%%%%%%%%%%%%%%%%%%%%
%%%%%%%%%%%%%%%%%%%%%%%%%%%%%%%%%%%%%%%%%%%%%%%%%
\begin{equation}\label{eq:limthetainfty}
\lim_{\theta\to\infty}\mu=\frac{\alpha}{\MyGammaSIMO},
\end{equation}
where $\alpha$ is a constant. Using \cite[eq. 3.381-4]{Ryzhik}, this constant is obtained as
%%%%%%%%%%%%%%%%%%%%%%%%%%%%%%%%%%%%%%%%%%%%%%%%%
%%%%%%%%%%%%%%%%%%%%%%%%%%%%%%%%%%%%%%%%%%%%%%%%%
%%% EQUATION 31 %%%%%%%%%%%%%%%%%%%%%%%%%%%%%%%%%
%%%%%%%%%%%%%%%%%%%%%%%%%%%%%%%%%%%%%%%%%%%%%%%%%
%%%%%%%%%%%%%%%%%%%%%%%%%%%%%%%%%%%%%%%%%%%%%%%%%
\begin{align}
\alpha=&\left(\frac{M_t}{\Mygammazero^{M_r}\Gamma(M_r)}\sum_{m=0}^{M_t-1}{M_t-1\choose m}(-1)^m\times\right.\nonumber\\
&\left.\sum_{q=0}^{Q(m)}c_q^{(m)}\left(\frac{m+1}{\Mygammazero}\right)^{-(M_r+q-1)}\Gamma(M_r+q-1)\right)^{-1}.
\end{align}
Here, $\mu\MyGammaSIMO$ converges to the constant $\alpha$, and therefore the effective capacity is
%%%%%%%%%%%%%%%%%%%%%%%%%%%%%%%%%%%%%%%%%%%%%%%%%
%%%%%%%%%%%%%%%%%%%%%%%%%%%%%%%%%%%%%%%%%%%%%%%%%
%%% EQUATION 32 %%%%%%%%%%%%%%%%%%%%%%%%%%%%%%%%%
%%%%%%%%%%%%%%%%%%%%%%%%%%%%%%%%%%%%%%%%%%%%%%%%%
%%%%%%%%%%%%%%%%%%%%%%%%%%%%%%%%%%%%%%%%%%%%%%%%%
\begin{equation}
E_C^{(\infty)}=\lim_{\theta\to\infty}E_C(\theta)=BT\log_2(1+\alpha)
\end{equation}
which reveals a constant rate when very high QoS is required.
%%%%%%%%%%%%%%%%%%%%%%%%%%%%%%%%%%%%%%%%%%%%%%%%%
%%%%%%%%%%%%%%%%%%%%%%%%%%%%%%%%%%%%%%%%%%%%%%%%%
%%% SECTION 5 %%%%%%%%%%%%%%%%%%%%%%%%%%%%%%%%%%%
%%%%%%%%%%%%%%%%%%%%%%%%%%%%%%%%%%%%%%%%%%%%%%%%%
%%%%%%%%%%%%%%%%%%%%%%%%%%%%%%%%%%%%%%%%%%%%%%%%%
\section{Simulation Results}\label{sec:5}
In all simulations, we assume $B=100\textrm{KHz}$,~$T=1\textrm{msec}$, and in each average SNR, the Mont-Carlo simulation is repeated $1,000,000$ times. Note that for simple representation, the normalized effective capacity $\overline{E_C}(\theta)=E_C(\theta)/(BT)$ is plotted in the following figures.

%%%%%%%%%%%%%%%%%%%%%%%%%%%%%%%%%%%%%%%%%%%%%%%%%
%%%%%%%%%%%%%%%%%%%%%%%%%%%%%%%%%%%%%%%%%%%%%%%%%
%%% FIGURE 2 %%%%%%%%%%%%%%%%%%%%%%%%%%%%%%%%%%%%
%%%%%%%%%%%%%%%%%%%%%%%%%%%%%%%%%%%%%%%%%%%%%%%%%
%%%%%%%%%%%%%%%%%%%%%%%%%%%%%%%%%%%%%%%%%%%%%%%%%
\begin{figure}
\begin{center}
\includegraphics[width=\linewidth]{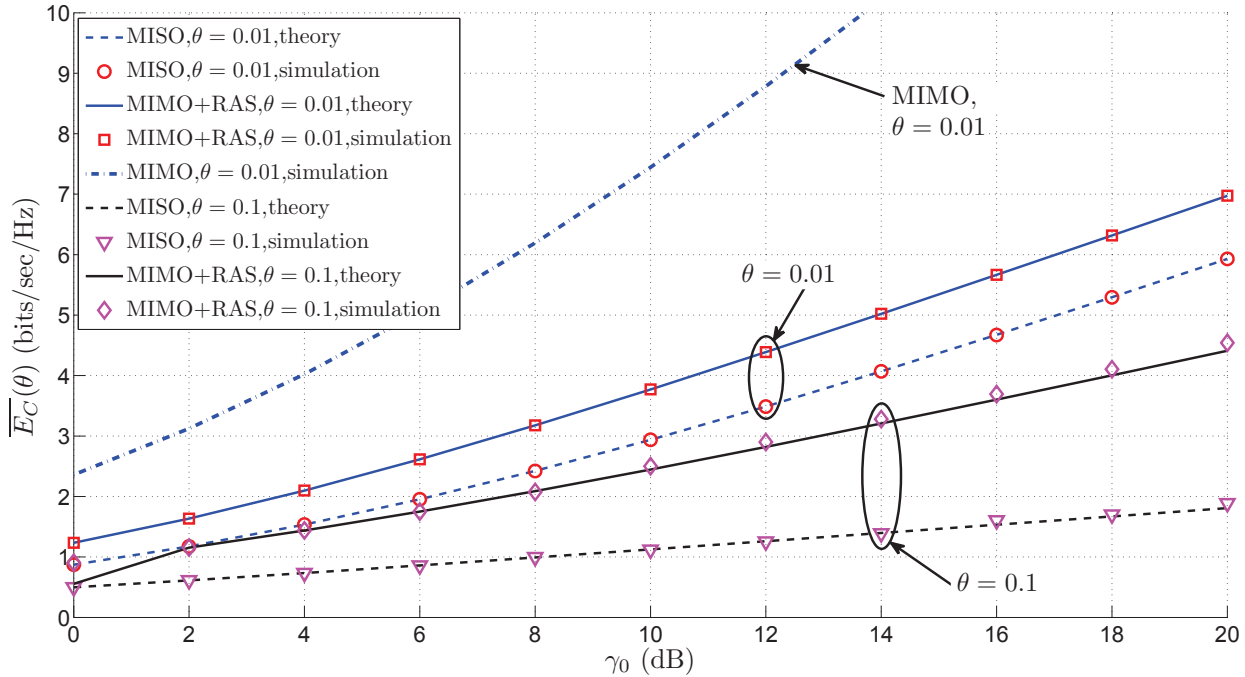}
\end{center}
\caption{Normalized effective capacity versus the average SNR in the $3\times3$ MIMO system with RAS and in the $3\times1$ MISO system.}
\label{fig:2}
\end{figure}
The normalized effective capacity with the RAS scheme is plotted versus $\Mygammazero$ in Fig.~\ref{fig:2} where $\theta=0.01$ and $\theta=0.1$. For more comparison, the normalized effective capacity of a $3\times1$ MISO system ($\theta=0.01$ and $\theta=0.1$) and a $3\times3$ MIMO system ($\theta=0.01$) is also plotted. Tight agreement between theory and simulation is clear here. The advantage of the RAS is observed in this figure where we have near $1~\textrm{bits/sec/Hz}$ and $2~\textrm{bits/sec/Hz}$ effective capacity gains compared with the equivalent MISO systems for $\theta=0.01$ and $\theta=0.1$ respectively. Therefore, we can suggest RAS technique specially when we need high QoS in the network. In addition, we observe that when the QoS exponent $\theta$ increases and more strict QoS is required, the effective capacity decreases.

%%%%%%%%%%%%%%%%%%%%%%%%%%%%%%%%%%%%%%%%%%%%%%%%%
%%%%%%%%%%%%%%%%%%%%%%%%%%%%%%%%%%%%%%%%%%%%%%%%%
%%% FIGURE 3 %%%%%%%%%%%%%%%%%%%%%%%%%%%%%%%%%%%%
%%%%%%%%%%%%%%%%%%%%%%%%%%%%%%%%%%%%%%%%%%%%%%%%%
%%%%%%%%%%%%%%%%%%%%%%%%%%%%%%%%%%%%%%%%%%%%%%%%%
\begin{figure}
\begin{center}
\includegraphics[width=\linewidth]{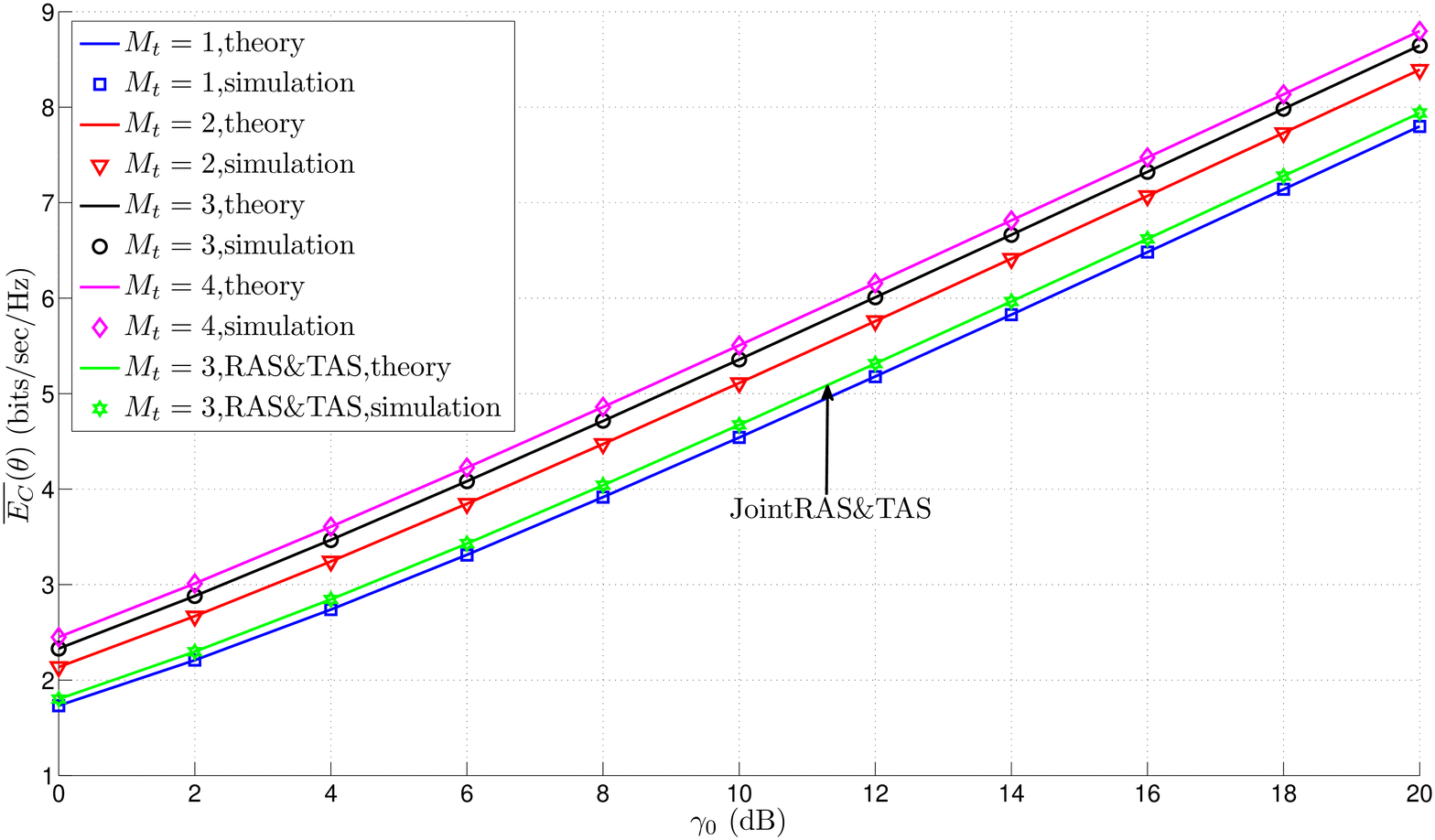}
\end{center}
\caption{Normalized effective capacity versus the average SNR and $\theta=0.01$ when $M_t$ varies from $1$ up to $4$ and $M_r=3$.}
\label{fig:3}
\end{figure}
Next we show that the effective capacity increases when the number of transmit antenna increases. For this purpose we plot the normalized effective capacity versus $\Mygammazero$ and $\theta=0.01$ in Fig.~\ref{fig:3}. We assume TAS scheme with the optimal power adaptation where $M_r=3$ and $M_t$ varies from $1$ up to $4$. The performance of a $3\times3$ MIMO system with joint receive and transmit AS is also presented here. The gains diminish with the increase in the number of transmit antennas. Therefore, it seems that using a MIMO system with two antennas at the transmitter with the TAS scheme has more advantages in contrast to other structures. The effective capacity in a MIMO system with the joint receive and transmit AS is more than the effective capacity of a $1\times3$ system but not as much as the others. Consequently, MIMO system with the TAS scheme in addition to the receive diversity performs better than the MIMO system with the joint receive and transmit AS.

%%%%%%%%%%%%%%%%%%%%%%%%%%%%%%%%%%%%%%%%%%%%%%%%%
%%%%%%%%%%%%%%%%%%%%%%%%%%%%%%%%%%%%%%%%%%%%%%%%%
%%% FIGURE 4 %%%%%%%%%%%%%%%%%%%%%%%%%%%%%%%%%%%%
%%%%%%%%%%%%%%%%%%%%%%%%%%%%%%%%%%%%%%%%%%%%%%%%%
%%%%%%%%%%%%%%%%%%%%%%%%%%%%%%%%%%%%%%%%%%%%%%%%%
\begin{figure}
  \begin{subfigure}[b]{\linewidth}
                \centering
                \includegraphics[width=\linewidth]{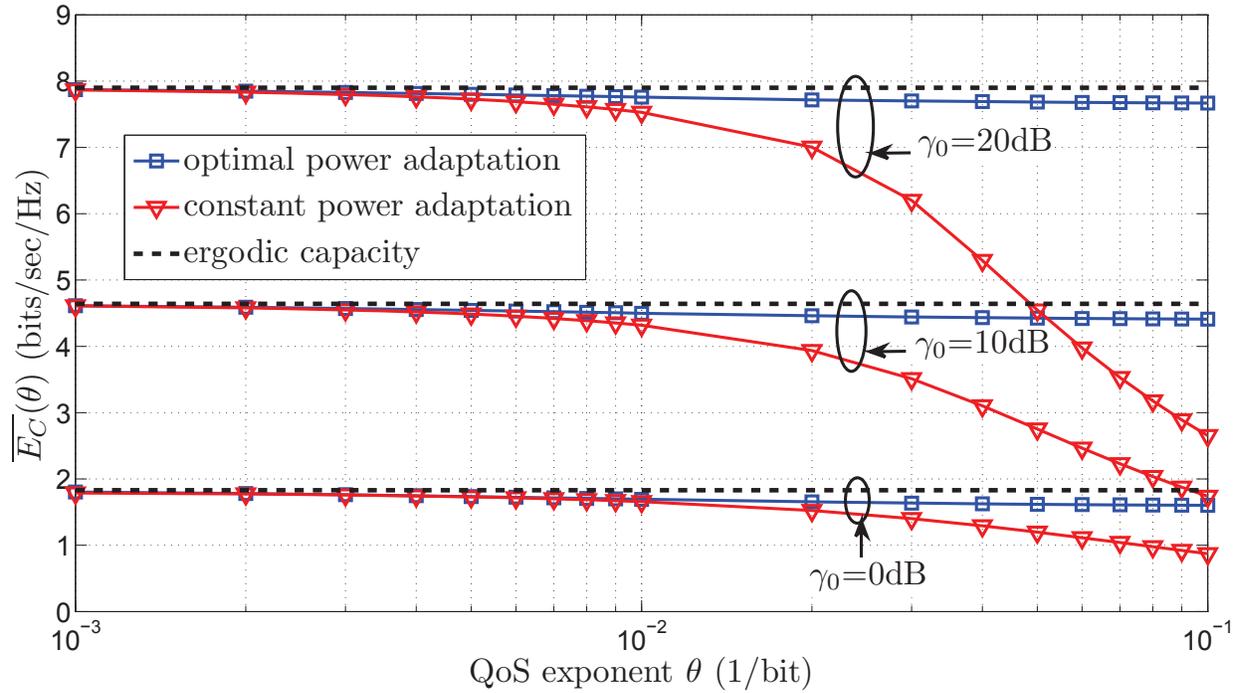}
                \caption{}
  \end{subfigure}\\
  \begin{subfigure}[b]{\linewidth}
                \centering
                \includegraphics[width=\linewidth]{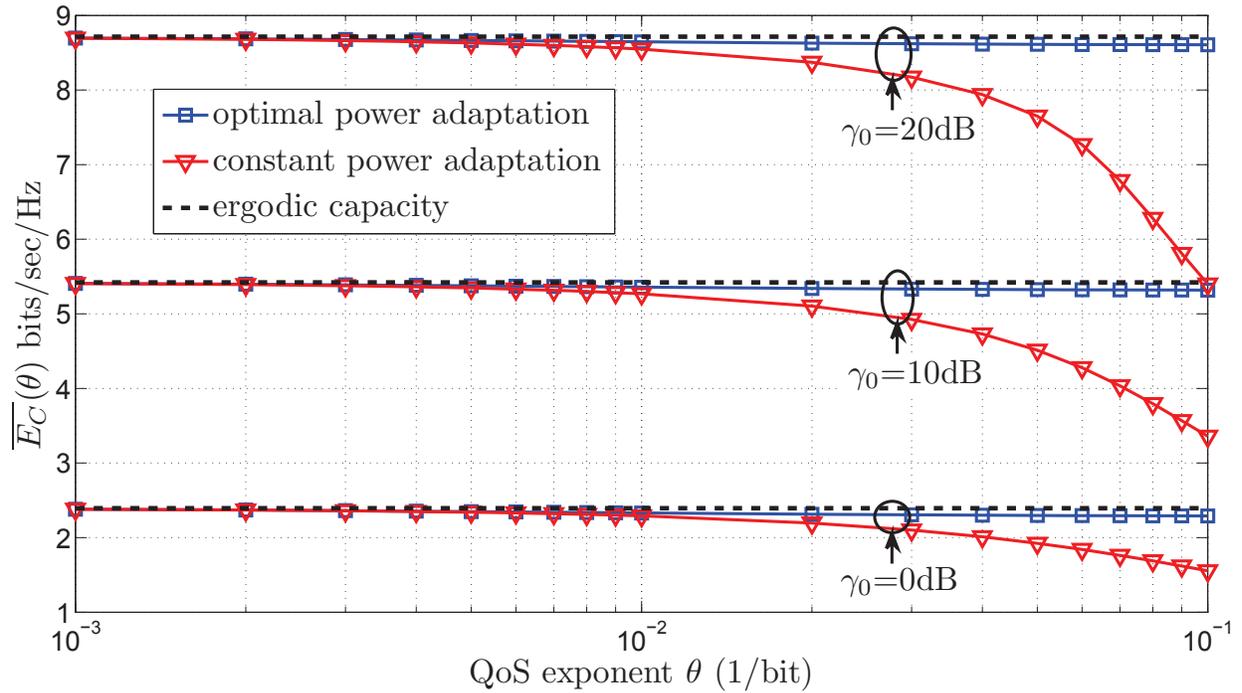}
                \caption{}
  \end{subfigure}
  \caption{Normalized effective capacity versus the QoS exponent $\theta$ with the optimal and constant power adaptation and normalized ergodic capacity in a (a) $2\times2$ and (b) $3\times3$ MIMO system with the TAS scheme.}
  \label{fig:4}
\end{figure}
The normalized effective capacity versus the QoS exponent $\theta$ is presented in Fig.~\ref{fig:4}. The Mont-Carlo simulations have a fine match with the theory, but for more obvious presentation, the theoretical results are plotted here. We assume a $2\times2$ and $3\times3$ MIMO system with the TAS scheme in three different average SNRs: $0\textrm{dB}$, $10\textrm{dB}$ and $20\textrm{dB}$. The normalized effective capacity with the optimal and constant power adaptation is compared in this figure where the gains of $2.67~\textrm{bits/sec/Hz}$ and $1.96~\textrm{bits/sec/Hz}$ at $\Mygammazero=10\textrm{dB}$ and $\theta=0.1$ are obtained in the $2\times2$ and $3\times3$ systems, respectively. Consequently, the optimal power adaptation is recommended, specifically when we have stringent QoS constraint. It would result in considerable improvement in the whole performance of the system. Moreover, the gap between the optimal and constant power adaptation becomes more distinct at the higher SNRs.

In Section \ref{subsec:C} the effective capacity is studied in two asymptotic cases. In the loose QoS requirement when $\theta\to 0$, the effective capacity with the optimal and constant power adaptation converge and reach to the ergodic capacity. Therefore, employing the optimal power coefficient $\mu$ does not have an advantage here. For more comparison, the normalized ergodic capacity $E_C^{(0)}/(BT)$ is also plotted in Fig.~\ref{fig:4} where the convergence of the effective capacity to that is clear.

%%%%%%%%%%%%%%%%%%%%%%%%%%%%%%%%%%%%%%%%%%%%%%%%%
%%%%%%%%%%%%%%%%%%%%%%%%%%%%%%%%%%%%%%%%%%%%%%%%%
%%% FIGURE 5 %%%%%%%%%%%%%%%%%%%%%%%%%%%%%%%%%%%%
%%%%%%%%%%%%%%%%%%%%%%%%%%%%%%%%%%%%%%%%%%%%%%%%%
%%%%%%%%%%%%%%%%%%%%%%%%%%%%%%%%%%%%%%%%%%%%%%%%%
\begin{figure}
  \begin{subfigure}[h]{\linewidth}
                \centering
                \includegraphics[width=\linewidth]{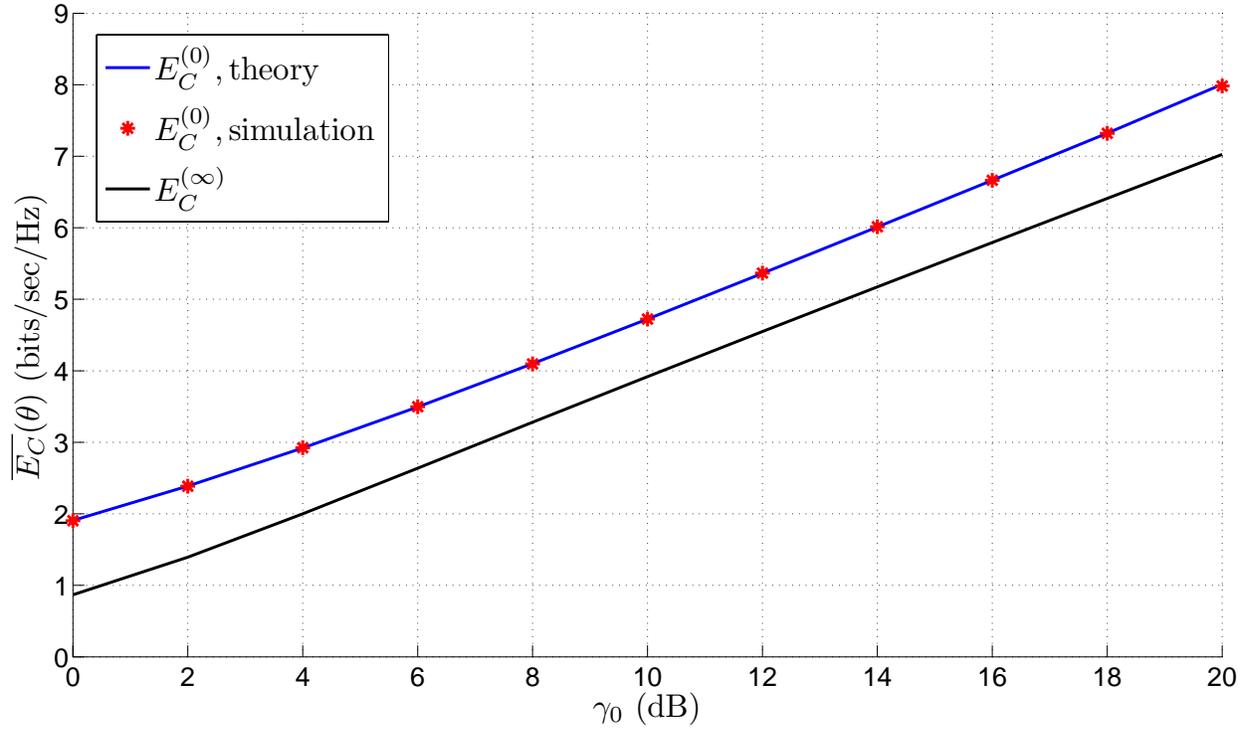}
                \caption{}
  \end{subfigure}\\
  \begin{subfigure}[h]{\linewidth}
                \centering
                \includegraphics[width=\linewidth]{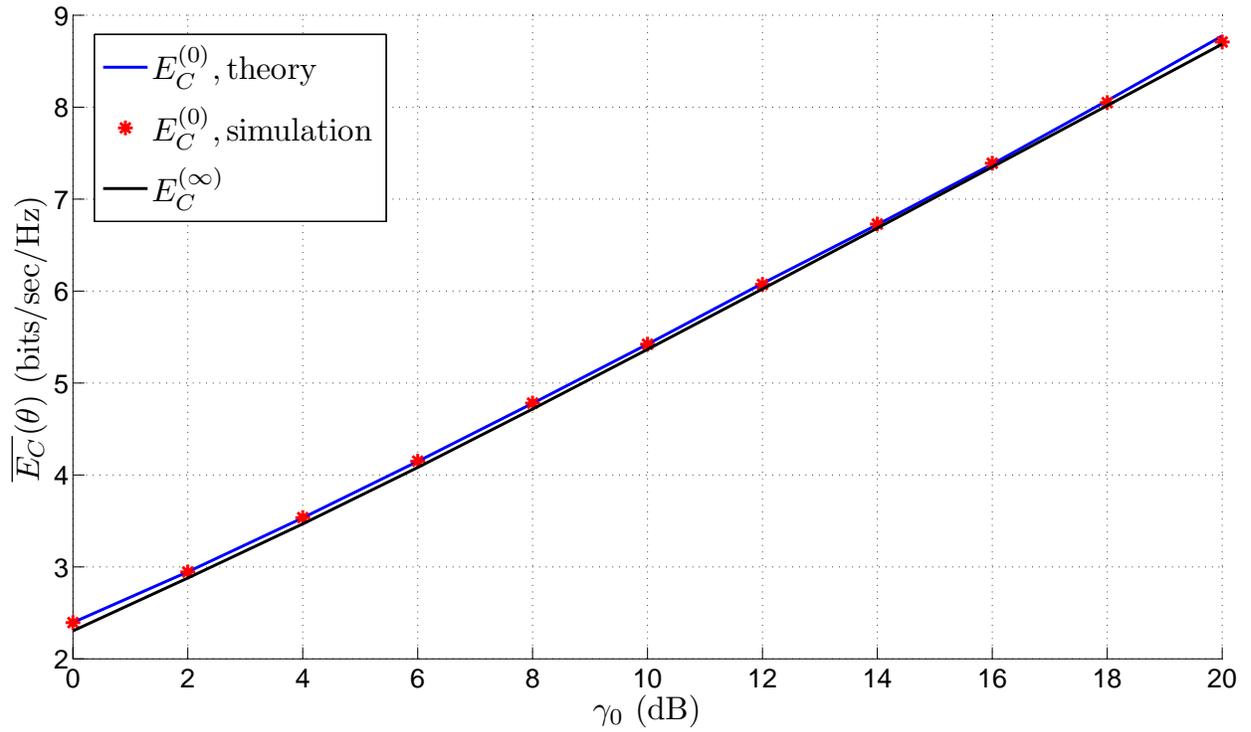}
                \caption{}
  \end{subfigure}
  \caption{Normalized asymptotic effective capacity performance versus the average SNR in (a) $1\times3$ SIMO without AS and (b) $3\times3$ MIMO with the TAS scheme.}
  \label{fig:5}
\end{figure}
We studied the the effective capacity performance in two asymptotic cases. In Fig.~\ref{fig:5}, $E_C^{(0)}/(BT)$ (theory and simulation) and $E_C^{(\infty)}/(BT)$ (theory) are plotted versus $\Mygammazero$ in a $1\times3$ SIMO system without AS and $3\times3$ MIMO systems with the TAS scheme. The optimal power adaptation is also assumed for these systems. To validate the closed-form expression which has been derived for $E_C^{(0)}$ (ergodic capacity), the theory and simulation results for $E_C^{(0)}$ are compared here where both results verify each other. In Fig.~\ref{fig:5} when $M_t$ increases the two asymptotic rates come closer. Subsequently we can conclude that the TAS scheme can keep the effective capacity near the ergodic capacity in all situations even at the high QoS constraint. Since the ergodic capacity determines the upper bound on the capacity rate, this result would be so elegant.
%%%%%%%%%%%%%%%%%%%%%%%%%%%%%%%%%%%%%%%%%%%%%%%%%
%%%%%%%%%%%%%%%%%%%%%%%%%%%%%%%%%%%%%%%%%%%%%%%%%
%%% SECTION 6 %%%%%%%%%%%%%%%%%%%%%%%%%%%%%%%%%%%
%%%%%%%%%%%%%%%%%%%%%%%%%%%%%%%%%%%%%%%%%%%%%%%%%
%%%%%%%%%%%%%%%%%%%%%%%%%%%%%%%%%%%%%%%%%%%%%%%%%
\section{Conclusion}\label{sec:6}
Effective capacity is an interesting topic and study of this subject in a practical MIMO system with RAS or TAS scheme is considered here. In the MIMO systems when there is no CSI at the transmitter, we adopt RAS technique and a closed-form solution for the maximum constant arrival rate with the QoS guarantee is extracted. In this case the effective capacity is compared with the performance of the equivalent MISO systems. We show that the improvements in the MIMO systems with RAS increases when the QoS exponent $\theta$ is increased. On the other hand in the MIMO systems with available CSI at the transmitter, TAS scheme in addition to an optimal power adaptation is employed. A closed-form solution for the optimal power and the effective capacity is also derived here and the advantages of this optimal power allocation in the effective capacity rate are also displayed. Then, the optimal power and the effective capacity are examined in two asymptotic cases when the QoS exponent $\theta$ tends to zero or infinity. We indicate that in the large and moderate value of SNRs when the QoS exponent $\theta$ approaches to zero, the optimal power converges to a constant value.

In addition, in a usual circumstance when the QoS exponent $\theta$ increases, the effective capacity decreases. We show that the TAS scheme can keep the effective capacity near the ergodic capacity even at large value of $\theta$. Therefore, when high QoS is required, TAS scheme is strictly suggested.
%%%%%%%%%%%%%%%%%%%%%%%%%%%%%%%%%%%%%%%%%%%%%%%%%
%%%%%%%%%%%%%%%%%%%%%%%%%%%%%%%%%%%%%%%%%%%%%%%%%
%%% APPENDIX A %%%%%%%%%%%%%%%%%%%%%%%%%%%%%%%%%%
%%%%%%%%%%%%%%%%%%%%%%%%%%%%%%%%%%%%%%%%%%%%%%%%%
%%%%%%%%%%%%%%%%%%%%%%%%%%%%%%%%%%%%%%%%%%%%%%%%%
\appendices
\section{Convergence of the Optimal Power Coefficient to a Constant}\label{appendix:1}
In this part the convergence of the optimal power coefficient $\mu$ to the constant one is proved. Suppose that when $\Mygammazero\to\infty$ we have $\MyGammazero\to1$. Since $\MyGammazero$ has a finite value, we can simply prove that
%%%%%%%%%%%%%%%%%%%%%%%%%%%%%%%%%%%%%%%%%%%%%%%%%
%%%%%%%%%%%%%%%%%%%%%%%%%%%%%%%%%%%%%%%%%%%%%%%%%
%%% EQUATION 33 %%%%%%%%%%%%%%%%%%%%%%%%%%%%%%%%%
%%%%%%%%%%%%%%%%%%%%%%%%%%%%%%%%%%%%%%%%%%%%%%%%%
%%%%%%%%%%%%%%%%%%%%%%%%%%%%%%%%%%%%%%%%%%%%%%%%%
\begin{equation}\label{eq:33}
\lim_{\Mygammazero\to\infty}\mathbb{P}\left\{\MyGammaSIMO<\MyGammazero\right\}=\lim_{\Mygammazero\to\infty}\mathbb{P}\left\{\MyGammaSIMO<1\right\}=0
\end{equation}
where $\mathbb{P}$ represents the probability function. With this probabilistic result, we can conclude that $\MyGammaSIMO\gg1$. In addition when $\theta\to 0$, we use both (\ref{eq:limthetazero}) and (\ref{eq:33}) to obtain
%%%%%%%%%%%%%%%%%%%%%%%%%%%%%%%%%%%%%%%%%%%%%%%%%
%%%%%%%%%%%%%%%%%%%%%%%%%%%%%%%%%%%%%%%%%%%%%%%%%
%%% EQUATION 34 %%%%%%%%%%%%%%%%%%%%%%%%%%%%%%%%%
%%%%%%%%%%%%%%%%%%%%%%%%%%%%%%%%%%%%%%%%%%%%%%%%%
%%%%%%%%%%%%%%%%%%%%%%%%%%%%%%%%%%%%%%%%%%%%%%%%%
\begin{equation}\label{eq:applimthetazero1}
\lim_{\theta\to 0,\Mygammazero\to\infty}\mu=\frac{1}{\MyGammazero}-\frac{1}{\MyGammaSIMO}=\frac{\MyGammaSIMO-1}{\MyGammaSIMO}=1.
\end{equation}
where we assume that $\MyGammaSIMO\gg1$. The proof is complete here.

Note that we assumed $\MyGammazero=1$ when $\theta\to 0$ and $\Mygammazero\to\infty$. Now in the followings, we prove it in two different cases.
%%%%%%%%%%%%%%%%%%%%%%%%%%%%%%%%%%%%%%%%%%%%%%%%%
%%%%%%%%%%%%%%%%%%%%%%%%%%%%%%%%%%%%%%%%%%%%%%%%%
%%% SUBSECTION A %%%%%%%%%%%%%%%%%%%%%%%%%%%%%%%%
%%%%%%%%%%%%%%%%%%%%%%%%%%%%%%%%%%%%%%%%%%%%%%%%%
%%%%%%%%%%%%%%%%%%%%%%%%%%%%%%%%%%%%%%%%%%%%%%%%%
\subsection{$M_r=1,~2$ and Arbitrary $M_t$}\label{subsec:appA}
When $\theta\to0$, the asymptotic behavior of the average power constraint (\ref{eq:expextationmu}) is reduced to
%%%%%%%%%%%%%%%%%%%%%%%%%%%%%%%%%%%%%%%%%%%%%%%%%
%%%%%%%%%%%%%%%%%%%%%%%%%%%%%%%%%%%%%%%%%%%%%%%%%
%%% EQUATION 35 %%%%%%%%%%%%%%%%%%%%%%%%%%%%%%%%%
%%%%%%%%%%%%%%%%%%%%%%%%%%%%%%%%%%%%%%%%%%%%%%%%%
%%%%%%%%%%%%%%%%%%%%%%%%%%%%%%%%%%%%%%%%%%%%%%%%%
\begin{align}\label{eq:appexpectationmu1}
&\mathbb{E}\{\mu\}=\frac{M_t}{\Mygammazero^{M_r}\Gamma(M_r)}\sum_{m=0}^{M_t-1}{M_t-1\choose m}(-1)^m\sum_{q=0}^{Q(m)}c_q^{(m)}\times\nonumber\\
&\left[\frac{1}{\MyGammazero}\left(\frac{m+1}{\Mygammazero}\right)^{-(M_r+q)}\Gamma\left(M_r+q,\frac{(m+1)\MyGammazero}{\Mygammazero}\right)-\right.\nonumber\\
&\left.\left(\frac{m+1}{\Mygammazero}\right)^{-(M_r+q-1)}\Gamma\left(M_r+q-1,\frac{(m+1)\MyGammazero}{\Mygammazero}\right)\right]=1.
\end{align}
In addition when $\Mygammazero\to\infty$, we can rewrite (\ref{eq:appexpectationmu1}) as
%%%%%%%%%%%%%%%%%%%%%%%%%%%%%%%%%%%%%%%%%%%%%%%%%
%%%%%%%%%%%%%%%%%%%%%%%%%%%%%%%%%%%%%%%%%%%%%%%%%
%%% EQUATION 36 %%%%%%%%%%%%%%%%%%%%%%%%%%%%%%%%%
%%%%%%%%%%%%%%%%%%%%%%%%%%%%%%%%%%%%%%%%%%%%%%%%%
%%%%%%%%%%%%%%%%%%%%%%%%%%%%%%%%%%%%%%%%%%%%%%%%%
\begin{align}\label{eq:appexpectationmu2}
&\mathbb{E}\{\mu\}=\frac{M_t}{\Mygammazero^{M_r}\Gamma(M_r)}\sum_{m=0}^{M_t-1}{M_t-1\choose m}(-1)^m\sum_{q=0}^{Q(m)}c_q^{(m)}\times\nonumber\\
&\left[\frac{1}{\MyGammazero}\left(\frac{m+1}{\Mygammazero}\right)^{-(M_r+q)}\Gamma(M_r+q)\right]=1,
\end{align}
where $((m+1)/\Mygammazero)^{-(M_r+q-1)}$ is neglected against $((m+1)/\Mygammazero)^{-(M_r+q)}$, and the upper incomplete Gamma function goes to the Gamma function when its second argument tends to zero. We have defined $c_q^{(m)}$ in Section \ref{sec:4} as
%%%%%%%%%%%%%%%%%%%%%%%%%%%%%%%%%%%%%%%%%%%%%%%%%
%%%%%%%%%%%%%%%%%%%%%%%%%%%%%%%%%%%%%%%%%%%%%%%%%
%%% EQUATION 37 %%%%%%%%%%%%%%%%%%%%%%%%%%%%%%%%%
%%%%%%%%%%%%%%%%%%%%%%%%%%%%%%%%%%%%%%%%%%%%%%%%%
%%%%%%%%%%%%%%%%%%%%%%%%%%%%%%%%%%%%%%%%%%%%%%%%%
\begin{equation}\label{eq:37}
\left(\sum_{k=0}^{M_r-1}\frac{x^k}{\Mygammazero^kk!}\right)^m=\sum_{q=0}^{Q(m)}c_q^{(m)}x^q,~~~0\leq m\leq M_t-1.
\end{equation}
With $M_r=1$ and arbitrary number of transmit antenna $M_t$, we have $Q(m)=m(M_r-1)=0$, and therefore, $c_q^{(m)}=1$ for $q=0$ and $0\leq m\leq M_t-1$. In this case according to (\ref{eq:appexpectationmu2}) we can write
%%%%%%%%%%%%%%%%%%%%%%%%%%%%%%%%%%%%%%%%%%%%%%%%
%%%%%%%%%%%%%%%%%%%%%%%%%%%%%%%%%%%%%%%%%%%%%%%%%
%%% EQUATION 38 %%%%%%%%%%%%%%%%%%%%%%%%%%%%%%%%%
%%%%%%%%%%%%%%%%%%%%%%%%%%%%%%%%%%%%%%%%%%%%%%%%%
%%%%%%%%%%%%%%%%%%%%%%%%%%%%%%%%%%%%%%%%%%%%%%%%%
\begin{align}\label{eq:38}
\mathbb{E}\{\mu\}&=\frac{M_t}{\Mygammazero}\sum_{m=0}^{M_t-1}{M_t-1\choose m}(-1)^m\frac{1}{\MyGammazero}\left(\frac{m+1}{\Mygammazero}\right)^{-1}\nonumber\\
&=\frac{1}{\MyGammazero}\sum_{m=0}^{M_t-1}{M_t\choose m+1}(-1)^m\nonumber\\
&=\frac{1}{\MyGammazero}\left[-\sum_{\tilde{m}=1}^{M_t}{M_t\choose \tilde{m}}(-1)^{\tilde{m}}\right]\nonumber\\
&=\frac{1}{\MyGammazero}\left[-\sum_{\tilde{m}=0}^{M_t}{M_t\choose \tilde{m}}(-1)^{\tilde{m}}+1\right]=\frac{1}{\MyGammazero}
\end{align}
where $m+1$ is replaced by $\tilde{m}$. Since $\mathbb{E}\{\mu\}=1$, we can conclude that $\MyGammazero=1$. Moreover, for $M_r=2$ and arbitrary number of transmit antenna $M_t$, we have $Q(m)=m$, and from (\ref{eq:37}), we show
%%%%%%%%%%%%%%%%%%%%%%%%%%%%%%%%%%%%%%%%%%%%%%%%%
%%%%%%%%%%%%%%%%%%%%%%%%%%%%%%%%%%%%%%%%%%%%%%%%%
%%% EQUATION 39 %%%%%%%%%%%%%%%%%%%%%%%%%%%%%%%%%
%%%%%%%%%%%%%%%%%%%%%%%%%%%%%%%%%%%%%%%%%%%%%%%%%
%%%%%%%%%%%%%%%%%%%%%%%%%%%%%%%%%%%%%%%%%%%%%%%%%
\begin{equation}
\left(1+\frac{x}{\Mygammazero}\right)^m=\sum_{q=0}^{m}{m\choose q}\frac{x^q}{\Mygammazero^q},~~~0\leq m\leq M_t-1
\end{equation}
which leads to
%%%%%%%%%%%%%%%%%%%%%%%%%%%%%%%%%%%%%%%%%%%%%%%%%
%%%%%%%%%%%%%%%%%%%%%%%%%%%%%%%%%%%%%%%%%%%%%%%%%
%%% EQUATION 40 %%%%%%%%%%%%%%%%%%%%%%%%%%%%%%%%%
%%%%%%%%%%%%%%%%%%%%%%%%%%%%%%%%%%%%%%%%%%%%%%%%%
%%%%%%%%%%%%%%%%%%%%%%%%%%%%%%%%%%%%%%%%%%%%%%%%%
\begin{equation}
c_q^{(m)}={m\choose q}\frac{1}{\Mygammazero^q}
\end{equation}
for $0\leq q\leq m$ and $0\leq m\leq M_t-1$. Now, (\ref{eq:appexpectationmu2}) is reduced to
%%%%%%%%%%%%%%%%%%%%%%%%%%%%%%%%%%%%%%%%%%%%%%%%
%%%%%%%%%%%%%%%%%%%%%%%%%%%%%%%%%%%%%%%%%%%%%%%%%
%%% EQUATION 41 %%%%%%%%%%%%%%%%%%%%%%%%%%%%%%%%%
%%%%%%%%%%%%%%%%%%%%%%%%%%%%%%%%%%%%%%%%%%%%%%%%%
%%%%%%%%%%%%%%%%%%%%%%%%%%%%%%%%%%%%%%%%%%%%%%%%%
\begin{align}\label{eq:41}
&\mathbb{E}\{\mu\}=\frac{M_t}{\Mygammazero^2}\sum_{m=0}^{M_t-1}{M_t-1\choose m}(-1)^m\times\nonumber\\
&\left[\sum_{q=0}^{m}{m\choose q}\frac{1}{\Mygammazero^q\MyGammazero}\left(\frac{m+1}{\Mygammazero}\right)^{-(2+q)}\Gamma(2+q)\right]\nonumber\\
&=\frac{M_t}{\MyGammazero}\sum_{m=0}^{M_t-1}{M_t-1\choose m}(-1)^m\frac{1}{m+1}\sum_{q=0}^{m}{m\choose q}\frac{\Gamma(2+q)}{(m+1)^{1+q}}.
\end{align}
The inner sum is calculated first. If the summation is done backward one by one, it can be shown that after $p$ steps we have
%%%%%%%%%%%%%%%%%%%%%%%%%%%%%%%%%%%%%%%%%%%%%%%%
%%%%%%%%%%%%%%%%%%%%%%%%%%%%%%%%%%%%%%%%%%%%%%%%%
%%% EQUATION 42 %%%%%%%%%%%%%%%%%%%%%%%%%%%%%%%%%
%%%%%%%%%%%%%%%%%%%%%%%%%%%%%%%%%%%%%%%%%%%%%%%%%
%%%%%%%%%%%%%%%%%%%%%%%%%%%%%%%%%%%%%%%%%%%%%%%%%
\begin{align}\label{eq:42}
\sum_{q=m}^{m-p+1}&{m\choose q}\frac{\Gamma(2+q)}{(m+1)^{1+q}}=\nonumber\\
&\frac{\Gamma(m-p+3)}{(m+1)^{m-p+1}}\frac{m(m-1)(...)(m-p+3)}{(p-1)!},
\end{align}
and finally after $p=m+1$ steps, we have
%%%%%%%%%%%%%%%%%%%%%%%%%%%%%%%%%%%%%%%%%%%%%%%%
%%%%%%%%%%%%%%%%%%%%%%%%%%%%%%%%%%%%%%%%%%%%%%%%%
%%% EQUATION 43 %%%%%%%%%%%%%%%%%%%%%%%%%%%%%%%%%
%%%%%%%%%%%%%%%%%%%%%%%%%%%%%%%%%%%%%%%%%%%%%%%%%
%%%%%%%%%%%%%%%%%%%%%%%%%%%%%%%%%%%%%%%%%%%%%%%%%
\begin{align}\label{eq:43}
\sum_{q=m}^{0}{m\choose q}\frac{\Gamma(2+q)}{(m+1)^{1+q}}=\sum_{q=0}^{m}{m\choose q}\frac{\Gamma(2+q)}{(m+1)^{1+q}}=1.
\end{align}
Note that, (\ref{eq:42}) can be simply proved using mathematical induction. Substituting (\ref{eq:43}) in (\ref{eq:41}), we get
%%%%%%%%%%%%%%%%%%%%%%%%%%%%%%%%%%%%%%%%%%%%%%%%
%%%%%%%%%%%%%%%%%%%%%%%%%%%%%%%%%%%%%%%%%%%%%%%%%
%%% EQUATION 44 %%%%%%%%%%%%%%%%%%%%%%%%%%%%%%%%%
%%%%%%%%%%%%%%%%%%%%%%%%%%%%%%%%%%%%%%%%%%%%%%%%%
%%%%%%%%%%%%%%%%%%%%%%%%%%%%%%%%%%%%%%%%%%%%%%%%%
\begin{align}\label{eq:44}
\mathbb{E}\{\mu\}=\frac{M_t}{\MyGammazero}\sum_{m=0}^{M_t-1}{M_t-1\choose m}(-1)^m\frac{1}{m+1}=\frac{1}{\MyGammazero},
\end{align}
where the same procedure according to (\ref{eq:38}) can be used to obtain $\MyGammazero=1$.
%%%%%%%%%%%%%%%%%%%%%%%%%%%%%%%%%%%%%%%%%%%%%%%%%
%%%%%%%%%%%%%%%%%%%%%%%%%%%%%%%%%%%%%%%%%%%%%%%%%
%%% SUBSECTION B %%%%%%%%%%%%%%%%%%%%%%%%%%%%%%%%
%%%%%%%%%%%%%%%%%%%%%%%%%%%%%%%%%%%%%%%%%%%%%%%%%
%%%%%%%%%%%%%%%%%%%%%%%%%%%%%%%%%%%%%%%%%%%%%%%%%
\subsection{$M_t=1,~2$ and Arbitrary $M_r$}\label{subsec:appB}
In this case $M_t=1$ and arbitrary $M_r$ receive antennas are available. Therefore $c_q^{(m)}=1$ for $q=0$ and $m=0$ and through (\ref{eq:appexpectationmu2}) we obtain
%%%%%%%%%%%%%%%%%%%%%%%%%%%%%%%%%%%%%%%%%%%%%%%%
%%%%%%%%%%%%%%%%%%%%%%%%%%%%%%%%%%%%%%%%%%%%%%%%%
%%% EQUATION 45 %%%%%%%%%%%%%%%%%%%%%%%%%%%%%%%%%
%%%%%%%%%%%%%%%%%%%%%%%%%%%%%%%%%%%%%%%%%%%%%%%%%
%%%%%%%%%%%%%%%%%%%%%%%%%%%%%%%%%%%%%%%%%%%%%%%%%
\begin{align}\label{eq:45}
\mathbb{E}\{\mu\}=\frac{1}{\Mygammazero^{M_r}\Gamma(M_r)}\frac{1}{\MyGammazero}\left(\frac{1}{\Mygammazero}\right)^{-M_r}\Gamma(M_r)=\frac{1}{\MyGammazero}=1,
\end{align}
therefore, $\MyGammazero=1$. In addition for $M_t=2$ and arbitrary number of receive antenna $M_r$, (\ref{eq:appexpectationmu2}) is reduced to
%%%%%%%%%%%%%%%%%%%%%%%%%%%%%%%%%%%%%%%%%%%%%%%%
%%%%%%%%%%%%%%%%%%%%%%%%%%%%%%%%%%%%%%%%%%%%%%%%%
%%% EQUATION 46 %%%%%%%%%%%%%%%%%%%%%%%%%%%%%%%%%
%%%%%%%%%%%%%%%%%%%%%%%%%%%%%%%%%%%%%%%%%%%%%%%%%
%%%%%%%%%%%%%%%%%%%%%%%%%%%%%%%%%%%%%%%%%%%%%%%%%
\begin{align}\label{eq:46}
\mathbb{E}\{\mu\}&=\frac{2}{\Mygammazero^{M_r}\Gamma(M_r)\MyGammazero}\left(\frac{1}{\Mygammazero}\right)^{-M_r}\Gamma(M_r)\nonumber\\
&-\frac{2}{\Mygammazero^{M_r}\Gamma(M_r)}\sum_{q=0}^{M_r-1}\frac{1}{\Mygammazero^qq!\MyGammazero}\left(\frac{2}{\Mygammazero}\right)^{-(M_r+q)}\Gamma(M_r+q)\nonumber\\
&=\frac{2}{\MyGammazero}-\frac{2}{\MyGammazero\Gamma(M_r)2^{M_r}}\sum_{q=0}^{M_r-1}\frac{\Gamma(M_r+q)}{q!2^q}\nonumber\\
&=\frac{2}{\MyGammazero}-\frac{2}{\MyGammazero2^{M_r-1}2^{M_r}}\sum_{q=0}^{M_r-1}{M_r-1+q\choose q}2^{M_r-1-q.}
\end{align}
To obtain a closed-form solution for the sum in (\ref{eq:46}), we can use \cite[eq. 0.241-10]{Ryzhik}
%%%%%%%%%%%%%%%%%%%%%%%%%%%%%%%%%%%%%%%%%%%%%%%%
%%%%%%%%%%%%%%%%%%%%%%%%%%%%%%%%%%%%%%%%%%%%%%%%%
%%% EQUATION 47 %%%%%%%%%%%%%%%%%%%%%%%%%%%%%%%%%
%%%%%%%%%%%%%%%%%%%%%%%%%%%%%%%%%%%%%%%%%%%%%%%%%
%%%%%%%%%%%%%%%%%%%%%%%%%%%%%%%%%%%%%%%%%%%%%%%%%
\begin{equation}\label{eq:47}
\sum_{i=0}^{I}{I+i\choose i}2^{I-i}=4^I.
\end{equation}
Therefore we have
%%%%%%%%%%%%%%%%%%%%%%%%%%%%%%%%%%%%%%%%%%%%%%%%
%%%%%%%%%%%%%%%%%%%%%%%%%%%%%%%%%%%%%%%%%%%%%%%%%
%%% EQUATION 48 %%%%%%%%%%%%%%%%%%%%%%%%%%%%%%%%%
%%%%%%%%%%%%%%%%%%%%%%%%%%%%%%%%%%%%%%%%%%%%%%%%%
%%%%%%%%%%%%%%%%%%%%%%%%%%%%%%%%%%%%%%%%%%%%%%%%%
\begin{equation}\label{eq:48}
\mathbb{E}\{\mu\}=\frac{2}{\MyGammazero}-\frac{2}{\MyGammazero2^{M_r-1}2^{M_r}}4^{M_r-1}=\frac{1}{\MyGammazero}=1.
\end{equation}
which indicate $\MyGammazero=1$ and finally the proof is concluded.
%%%%%%%%%%%%%%%%%%%%%%%%%%%%%%%%%%%%%%%%%%%%%%%%%
%%%%%%%%%%%%%%%%%%%%%%%%%%%%%%%%%%%%%%%%%%%%%%%%%
%%% APPENDIX B %%%%%%%%%%%%%%%%%%%%%%%%%%%%%%%%%%
%%%%%%%%%%%%%%%%%%%%%%%%%%%%%%%%%%%%%%%%%%%%%%%%%
%%%%%%%%%%%%%%%%%%%%%%%%%%%%%%%%%%%%%%%%%%%%%%%%%
\section{Ergodic Capacity}\label{appendix:2}
First we prove that the ergodic capacity can be achieved from the effective capacity when $\theta\to 0$. Using the Taylor expansion we have
%%%%%%%%%%%%%%%%%%%%%%%%%%%%%%%%%%%%%%%%%%%%%%%%%
%%%%%%%%%%%%%%%%%%%%%%%%%%%%%%%%%%%%%%%%%%%%%%%%%
%%% EQUATION 49 %%%%%%%%%%%%%%%%%%%%%%%%%%%%%%%%%
%%%%%%%%%%%%%%%%%%%%%%%%%%%%%%%%%%%%%%%%%%%%%%%%%
%%%%%%%%%%%%%%%%%%%%%%%%%%%%%%%%%%%%%%%%%%%%%%%%%
\begin{align}\label{eq:49}
E_C^{(0)}&=\lim_{\theta\to0}-\frac{1}{\theta}\ln\left(\mathbb{E}\left\{\left(1+\mu\MyGammaSIMO\right)^{-\Mythetat}\right\}\right)\nonumber\\
&=\lim_{\theta\to0}-\frac{1}{\theta}\ln\left(\mathbb{E}\left\{1-\Mythetat\ln\left(1+\mu\MyGammaSIMO\right)\right\}\right)\nonumber\\
&=\lim_{\theta\to0}-\frac{1}{\theta}\ln\left(1-\Mythetat\mathbb{E}\left\{\ln\left(1+\mu\MyGammaSIMO\right)\right\}\right)\nonumber\\
&=\frac{-\Mythetat\mathbb{E}\left\{\ln\left(1+\mu\MyGammaSIMO\right)\right\}}{-\theta}\nonumber\\
&=BT\mathbb{E}\left\{\log_2\left(1+\mu\MyGammaSIMO\right)\right\}
\end{align}
where $E_C^{(0)}$ denotes the ergodic capacity. Next, an exact closed-form solution for the ergodic capacity is extracted. By substituting (\ref{eq:limthetazero}) into (\ref{eq:49}) and using the given PDF for $\MyGammaSIMO$ we have
%%%%%%%%%%%%%%%%%%%%%%%%%%%%%%%%%%%%%%%%%%%%%%%%%
%%%%%%%%%%%%%%%%%%%%%%%%%%%%%%%%%%%%%%%%%%%%%%%%%
%%% EQUATION 50 %%%%%%%%%%%%%%%%%%%%%%%%%%%%%%%%%
%%%%%%%%%%%%%%%%%%%%%%%%%%%%%%%%%%%%%%%%%%%%%%%%%
%%%%%%%%%%%%%%%%%%%%%%%%%%%%%%%%%%%%%%%%%%%%%%%%%
\begin{align}\label{eq:50}
E_C^{(0)}=\frac{BT}{\ln2}\frac{M_t}{\Mygammazero^{M_r}\Gamma(M_r)}\sum_{m=0}^{M_t-1}{M_t-1\choose m}(-1)^m\sum_{q=0}^{Q(m)}c_q^{(m)}I_{1,q}
\end{align}
where
%%%%%%%%%%%%%%%%%%%%%%%%%%%%%%%%%%%%%%%%%%%%%%%%%
%%%%%%%%%%%%%%%%%%%%%%%%%%%%%%%%%%%%%%%%%%%%%%%%%
%%% EQUATION 51 %%%%%%%%%%%%%%%%%%%%%%%%%%%%%%%%%
%%%%%%%%%%%%%%%%%%%%%%%%%%%%%%%%%%%%%%%%%%%%%%%%%
%%%%%%%%%%%%%%%%%%%%%%%%%%%%%%%%%%%%%%%%%%%%%%%%%
\begin{align}\label{eq:51}
I_{1,q}=\int_{\MyGammazero}^{\infty}\ln\left(\frac{x}{\MyGammazero}\right)x^{q+M_r-q}e^{-(m+1)x/\Mygammazero}dx.
\end{align}
We can change the variable in (\ref{eq:51}) as $y=(x/\MyGammazero)-1$ and write $I_{1,q}$ as
%%%%%%%%%%%%%%%%%%%%%%%%%%%%%%%%%%%%%%%%%%%%%%%%%
%%%%%%%%%%%%%%%%%%%%%%%%%%%%%%%%%%%%%%%%%%%%%%%%%
%%% EQUATION 52 %%%%%%%%%%%%%%%%%%%%%%%%%%%%%%%%%
%%%%%%%%%%%%%%%%%%%%%%%%%%%%%%%%%%%%%%%%%%%%%%%%%
%%%%%%%%%%%%%%%%%%%%%%%%%%%%%%%%%%%%%%%%%%%%%%%%%
\begin{align}\label{eq:52}
I_{1,q}=&\MyGammazero^{q+Mr}e^{-(m+1)\MyGammazero/\Mygammazero}\times\nonumber\\
&\underbrace{\int_{0}^{\infty}\ln(1+y)(1+y)^{q+M_r-1}e^{-\frac{(m+1)\MyGammazero}{\Mygammazero}y}dy}_{I_{2,q}}.
\end{align}
Here we can use binomial expansion for $(1+y)^{q+M_r-1}$ to simplify $I_{2,q}$ as
%%%%%%%%%%%%%%%%%%%%%%%%%%%%%%%%%%%%%%%%%%%%%%%%%
%%%%%%%%%%%%%%%%%%%%%%%%%%%%%%%%%%%%%%%%%%%%%%%%%
%%% EQUATION 53 %%%%%%%%%%%%%%%%%%%%%%%%%%%%%%%%%
%%%%%%%%%%%%%%%%%%%%%%%%%%%%%%%%%%%%%%%%%%%%%%%%%
%%%%%%%%%%%%%%%%%%%%%%%%%%%%%%%%%%%%%%%%%%%%%%%%%
\begin{equation}\label{eq:53}
I_{2,q}=\sum_{r=0}^{q+M_r-1}{q+M_r-1\choose r}\underbrace{\int_{0}^{\infty}\ln(1+y)y^re^{-\frac{(m+1)\MyGammazero}{\Mygammazero}y}dy}_{I_{3,r}}.
\end{equation}
Now we define $g_0=\Mygammazero/[(m+1)\MyGammazero]$ and change the variable $z=y/g_0$ in (\ref{eq:53}) to obtain
%%%%%%%%%%%%%%%%%%%%%%%%%%%%%%%%%%%%%%%%%%%%%%%%%
%%%%%%%%%%%%%%%%%%%%%%%%%%%%%%%%%%%%%%%%%%%%%%%%%
%%% EQUATION 54 %%%%%%%%%%%%%%%%%%%%%%%%%%%%%%%%%
%%%%%%%%%%%%%%%%%%%%%%%%%%%%%%%%%%%%%%%%%%%%%%%%%
%%%%%%%%%%%%%%%%%%%%%%%%%%%%%%%%%%%%%%%%%%%%%%%%%
\begin{equation}\label{eq:54}
I_{3,r}=g_0^{r+1}\underbrace{\int_{0}^{\infty}\ln(1+g_0z)z^re^{-z}dz}_{I_{4,r}}
\end{equation}
where $I_{4,r}$ has a closed-form solution as \cite[eq. 4.337-5]{Ryzhik}
%%%%%%%%%%%%%%%%%%%%%%%%%%%%%%%%%%%%%%%%%%%%%%%%%
%%%%%%%%%%%%%%%%%%%%%%%%%%%%%%%%%%%%%%%%%%%%%%%%%
%%% EQUATION 55 %%%%%%%%%%%%%%%%%%%%%%%%%%%%%%%%%
%%%%%%%%%%%%%%%%%%%%%%%%%%%%%%%%%%%%%%%%%%%%%%%%%
%%%%%%%%%%%%%%%%%%%%%%%%%%%%%%%%%%%%%%%%%%%%%%%%%
\begin{align}\label{eq:55}
I_{4,r}=\sum_{\xi=0}^{r}\frac{r!}{(r-\xi)!}&\left[\frac{(-1)^{r-\xi-1}}{g_0^{r-\xi}}e^{1/g_0}\textrm{Ei}\left(\frac{-1}{g_0}\right)+\right.\nonumber\\
&\left.\sum_{\nu=1}^{r-\xi}\frac{(-1)^{r-\xi-\nu}(\nu-1)!}{g_0^{r-\xi-\nu}}\right]
\end{align}
and $\textrm{Ei}(.)$ denotes the exponential integral function \cite[eq. 8.211-1]{Ryzhik}.
%%%%%%%%%%%%%%%%%%%%%%%%%%%%%%%%%%%%%%%%%%%%%%%%%
%%%%%%%%%%%%%%%%%%%%%%%%%%%%%%%%%%%%%%%%%%%%%%%%%
%%% SECTION ACKNOWLEDGMENTS %%%%%%%%%%%%%%%%%%%%%
%%%%%%%%%%%%%%%%%%%%%%%%%%%%%%%%%%%%%%%%%%%%%%%%%
%%%%%%%%%%%%%%%%%%%%%%%%%%%%%%%%%%%%%%%%%%%%%%%%%
\section*{Acknowledgments}
We acknowledge the support of the Research Institute for Information and Communication Technology, Tehran, Iran, under Contract No. T/18128/500 for its role in the development of this research.
%%%%%%%%%%%%%%%%%%%%%%%%%%%%%%%%%%%%%%%%%%%%%%%%%
%%%%%%%%%%%%%%%%%%%%%%%%%%%%%%%%%%%%%%%%%%%%%%%%%
%%% REFERENCES %%%%%%%%%%%%%%%%%%%%%%%%%%%%%%%%%%
%%%%%%%%%%%%%%%%%%%%%%%%%%%%%%%%%%%%%%%%%%%%%%%%%
%%%%%%%%%%%%%%%%%%%%%%%%%%%%%%%%%%%%%%%%%%%%%%%%%

\end{document}